\documentclass[a4paper]{amsart}
\usepackage{amsthm}
\usepackage{amssymb}
\theoremstyle{plain}
\newtheorem{thm}{Theorem}[subsection]
\newtheorem*{thm*}{Theorem}
\newtheorem*{G-R}{Gale--Ryser Theorem}

\newtheorem{cor}[thm]{Corollary}

\newtheorem{prob}{Problem}[subsection]
\newtheorem{conj}[thm]{Conjectur}
\newtheorem{claim}[thm]{Claim}
\theoremstyle{definition}
\newtheorem{Ex}{Example}[subsection]

\swapnumbers
\theoremstyle{remark}
\newtheorem{rem}[thm]{Remark}
\numberwithin{equation}{section}

\newcommand{\norm}[1]{\left\Vert#1\right\Vert}
\newcommand{\abs}[1]{\left\vert#1\right\vert}

\newcommand{\Real}{\mathbb R}

\newcommand{\Spec}{\operatorname{Spec}}
\newcommand{\Flag}{\operatorname{\mathcal{F}\ell}}
\newcommand{\Tr}{\operatorname{Tr}}
\newcommand{\rk}{\operatorname{rk}}
\newcommand{\SU}{\operatorname{SU}}
\newcommand{\diag}{\operatorname{diag}}
\newcommand{\height}{\operatorname{ht}}
\begin{document}

\title[Quantum Margins]
{Quantum marginal problem and representations of the symmetric group}%
\author{Alexander Klyachko}%
\address{Bilkent University \newline\indent Bilkent, 06533 Ankara Turkey}%
\email{klyachko@fen.bilkent.edu.tr}%


\dedicatory{To 60 anniversary of Alain~Lascoux}%

\begin{abstract}
 We discuss existence of mixed state
$\rho_{AB}$ of two (or multy-) component system ${
\mathcal{H}}_{AB}={
\mathcal{H}}_{A}\otimes {\mathcal{H}}_{B}$ with reduced density
matrices ${\rho_A}$, ${\rho_B}$ and given spectra
$\lambda_{AB},\lambda_A,\lambda_B$. We give a complete solution of
the problem in terms of linear inequalities on the spectra,
accompanied with extensive tables of marginal inequalities,
including arrays up to 4 qubits. In the second part of the paper
we pursue another approach based on reduction of the problem to
representation theory of the symmetric group.
\end{abstract}
\maketitle
\tableofcontents

\section{Introduction}

The quantum marginal problem is about relations between  spectrum
of mixed state $\rho_{AB}$ of two (or multi) component system
$\mathcal{H}_{AB}=\mathcal{H}_A\otimes \mathcal{H}_B$ and that of
reduced states $\rho_A$ and $\rho_B$. Relations of this type, for
example, impose certain restrictions on manipulations with qubits
in quantum information theory.
The problem can be stated in plain language as follows. Let
$M=[m_{ijk}]$ be cubic complex matrix and $H_1,H_2,H_3$ be Gram
matrices formed by Hermitian dot products of parallel slices of
$M$. We seek for
relations between spectra of these matrices.

We pursue two different approach to this problem. The first is
based on reduction
to general  Berenstein-Sjamaar theorem \cite{Ber-Sja} applied to
subgroup $\SU(\mathcal{H}_A)\times
SU(\mathcal{H}_B)\subset\SU(\mathcal{H}_A\otimes \mathcal{H}_B)$.
The relevant geometry and combinatorics are explained in section
\ref{MargIneq2}. Theorems \ref{cubicleThm} and \ref{ExtrEdge} give
a pretty explicit  ansatz for producing quantum marginal
inequalities. For systems of rank $\le 4$ they are given in
Appendix. It covers all system with few hundreds, rather then
thousands, marginal inequalities. An important case of an array of
qubits considered separately in section \ref{n_qubit}. Modulo a
``standard" conjecture the marginal inequalities can be produced
in a very straightforward way, see Theorem \ref{Mod_Qbt}.

The number of marginal inequalities increases drastically with
rank of the system. This makes the above solution
inefficient for systems of big rank. In section \ref{ReprTheor} we
develop another approach based on reduction of the marginal
problem to decomposition of tensor product of irreducible
representations of the symmetric group, see Theorem \ref{mainThm}.
This approach, for example, allows answer questions about maximal
eigenvalue of a mixed state with given margins, see Theorem
\ref{Max}, or its rank, see Theorem \ref{Rank}. However the main
point here is not in new results, but in new vision, based on
connections between apparently very remote subjects. In last
section \ref{Appl2Repr} we consider some applications of the above
correspondence to back to representation theory of the symmetric
group.

It is my pleasure to dedicate the paper  to Alain~Lascoux, who was
one of the inventors and main contributor to theory of
Schubert polynomials \cite{L-Sch} and gave the first known
combinatorial description of tensor product for a class of
irreducible representations of the symmetric group \cite{Las}.
Both topics turn out to be entangled in a surprising way with
quantum marginal problem.


I am appreciated to Sergey Bravyi and  Patrick Hayden for useful
discussion during my visit to the Institute for Quantum
Information in Caltech,
and to John Pereskill for
kind invitation. This work was supported in part by the National
Science Foundation under grant EIA-0086038 through the Institute
for Quantum Information at the California Institute of Technology.

The text was prepared as a talk at mini conference ``Turing Days
04: Classical \& Quantum Computing" in Istanbul Bilgi University,
29--30 May, 2004. By that reason it contains background
information which some readers may find redundant. I decided to
keep
it untouched to make the paper easily accessible both for
physicists and mathematicians.

 \section{Classical marginal problem} \label{class}
The classical marginal problem is about existence of a ``body" or
probability density $p_I(x_I):=p(x_1,x_2,\ldots,x_n)$ in $
\mathbb{R}^I$, $I=\{1,2,\ldots,n\}$ with given projections onto
some coordinate subspaces $\mathbb{R}^J\subset
\mathbb{R}^I,J\subset I$
$$p_J(x_J)=\int_{\mathbb{R}^{\bar{J}}}p_I(x_I)dx_{\bar{J}},\quad \bar{J}=I\backslash J$$
called {\it margins\,} of $p_I$. The problem has a long history,
see \cite{Kl02} and references therein. Here we give only few
relevant examples.

\subsubsection{\!\!}{\it Univariant margins $p_i(x_i)$ are always
compatible\,.} Indeed consider $x_i$ as independent variables and
define joint distribution by equation
$$p_I(x_I)=\prod_{i\in I}p_i(x_i).$$

\subsubsection{\!\!}{ The following inequality is necessary for compatibility of
bivariant margins}
$$\langle x_i|x_i\rangle+
\langle x_j| x_j\rangle+
\langle x_k| x_k\rangle+
2\langle x_i| x_j\rangle+ 2\langle x_j| x_k\rangle+ 2\langle x_k|
x_i\rangle\ge0.$$ Indeed LHS is equal to variance $\langle
x_i+x_j+x_k\mid x_i+x_j+x_k\rangle\ge0$ {\em provided} joint
distribution $p_{ijk}$ exists.

\subsubsection{\!\!} So called {\em Bell's ineqalities}\, in quantum mechanics are
just compatibility conditions for marginal distributions
corresponding to {\em commuting\,} observables \cite{Kl02}.

\subsubsection{\!\!} \label{Planar}{\em Discrete version of the marginal problem\,} is about
existence of, say, cubic matrix $p_{ijk}\ge0$ with given
projections onto its facets
$$a_{ij}=\sum_k p_{ijk},\quad b_{jk}=\sum_i p_{ijk},\quad c_{ki}=\sum_j
p_{ijk}.$$ Compatibility conditions for such projections are still
unknown (so called {\em Planar Transport Polytope Problem\,}, see
\cite{EKK84}).

\subsubsection{\!\!} \label{restr}{\em Restricted marginal problem\,} is about existence of a
matrix with prescribed content, e.g. $0-1$, and given margins. In
this case even univariant marginal problem becomes nontrivial. As
an example recall the following classical result.
\begin{thm*}[Gale \cite{Gale, G-R}, Ryser \cite{Ryser,G-R}]
Partitions $\lambda,\mu$ are margins of a rectangular $0-1$ matrix
iff $\lambda\prec \mu^t$.
\end{thm*}

Here marginal values
$\lambda_1\ge\lambda_2\ge\cdots\ge\lambda_m\ge0$ arranged in
decreasing order are treated as {\it Young diagram\,} $\lambda$
with $i$-th row of length $\lambda_i$, $\mu^t$ stands for
transpose diagram, and the {\em majorization\,} or {\it
dominance\,} order $\lambda\prec\mu^t$ is defined by inequalities
\begin{align*}\lambda_1&\le\mu^t_1\\
\lambda_1+\lambda_2&\le\mu^t_1+\mu^t_2\\
\lambda_1+\lambda_2+\lambda_3&\le\mu^t_1+\mu^t_2+\mu^t_3\\
\cdots&\cdots\cdots
\end{align*}
The number of $0-1$ matrices with margins $\lambda,\mu$ is equal
to the number of pairs of tableaux of conjugate shape and contents
$\lambda,\mu$, see \cite{Macdonald}.

 We'll deal with quantum version of this problem below.
\section{Quantum marginal problem}
\subsection{Quantum margins} A background of a quantum system $A$ is
Hilbert space $\mathcal{H}_A$ called {\it state space}. We'll
consider only {\it finite systems\,}, for which
$\dim\mathcal{H}_A<\infty$. Actual state of the system is
described by unit vector $\psi\in\mathcal{H}_A$ for {\em pure
state\,} or by non negative Hermitian operator
$\rho:\mathcal{H}_A\rightarrow\mathcal{H}_A,
\operatorname{Tr}\rho=1$, called {\em density matrix\,}, for {\em mixed state\,}. Pure state $\psi$
corresponds to projection operator $|\psi\rangle\langle\psi|$ onto
$\psi$. Hence
\begin{equation}
\rho=\mbox{pure}\Longleftrightarrow\operatorname{rk}\rho=1
\Longleftrightarrow\operatorname{Spec}\rho=(1,0,0\cdots,0).
\end{equation}
{\it An observable\,} is Hermitian operator
$X_A:\mathcal{H}_A\rightarrow\mathcal{H}_A$. Taking  measurement
of $X_A$
while system is in state $\rho$ produces random quantity
$x_A\in\Spec X_A$ implicitly determined by expectations
$$\langle f(x_A)\rangle_\rho=\Tr(\rho
f(X_A))=\langle\psi|f(X_A)|\psi\rangle$$ of arbitrary function
$f(x)$ on $\Spec X_A$. The second equation holds for pure state
$\psi$.

Superposition principle of quantum mechanics implies that state
space of composite system $AB$ splits into tensor product
$$\mathcal{H}_{AB}=\mathcal{H}_A\otimes\mathcal{H}_B$$
of state spaces of the components, as opposed to direct product
$P_{AB}=P_A\times P_B$ of configuration spaces in classical
mechanics. Density matrix of composite system can be written as
linear combination
\begin{equation}
\rho_{AB}=\sum_\alpha a_\alpha L_A^\alpha\otimes L_B^\alpha
\end{equation}
where $L_A^\alpha, L_B^\alpha$ are linear operators in
$\mathcal{H}_A,\mathcal{H}_B$ respectively. Its {\it reduced
matrices\,} or {\em marginal states\,} are defined by equations
\begin{align}
\rho_A&=\sum_\alpha
a_\alpha\Tr(L_B^\alpha)L_A^\alpha:=\Tr_B(\rho_{AB}),\\
\rho_B&=\sum_\alpha
a_\alpha\Tr(L_A^\alpha)L_B^\alpha:=\Tr_A(\rho_{AB}).
\end{align}
They can be characterized by the following property
\begin{equation}
\langle X_A\rangle_{\rho_{AB}}=\Tr(\rho_{AB}X_A)=\Tr(\rho_A
X_A)=\langle X_A\rangle_{\rho_A},\quad
X_A:\mathcal{H}_A\rightarrow\mathcal{H}_A,
\end{equation}
which tells that {\it observation of subsystem $A$  gives the same
results as if $A$ would be in reduced state
$\rho_A=\Tr_B(\rho_{AB})$.} This justifies the terminology.


Margins of state $\rho_I$ of multicomponent system
$$\mathcal{H}_I=\bigotimes_{i\in I}\mathcal{H}_i=
\mathcal{H}_J\otimes\mathcal{H}_{\bar{J}},\quad J\subset I,
\bar{J}=I\backslash J$$
are defined in a similar way: $\rho_J=\Tr_{\bar{J}}(\rho_I)$.

\begin{Ex} In tensor algebra the above reduction
$\rho_I\mapsto\rho_J,J\subset I$ is known as {\it contraction\,}.
Most mathematicians are familiar with this procedure from
differential geometry, where, say, Ricci curvature
$\operatorname{Ric}:\mathcal{T}\rightarrow\mathcal{T}$ is defined
as contraction of Riemann curvature
$R:\mathcal{T}\otimes\mathcal{T}\rightarrow\mathcal{T}\otimes\mathcal{T}$
(here  $\mathcal{T}$ stands for tangent bundle).
\end{Ex}
\begin{Ex} Let's identify pure state of two component system
$$\psi=\sum_{ij}\psi_{ij}\alpha_i\otimes\beta_j\in\mathcal{H}_A\otimes\mathcal{H}_B$$
with its matrix $[\psi_{ij}]$ in orthonormal bases
$\alpha_i,\beta_j$ of $\mathcal{H}_A,\mathcal{H}_B$. Then margins
of $\psi$ in respective bases are given by matrices
\begin{equation}\rho_A=\psi^\dag\psi,\quad\rho_B=\psi\psi^\dag.\label{2-comp}\end{equation}
In striking difference with classical case margins of a pure
quantum state are mixed ones (provided
$\psi\neq\psi_A\otimes\psi_B$).
\end{Ex}
\begin{Ex} \label{slice}
A similar description holds  for multicomponent systems. For
example, write orthonormal components of tensor
$\psi\in\mathcal{H}_A\otimes\mathcal{H}_B\otimes\mathcal{H}_C$
into a cubic matrix $[\psi_{ijk}]$. Then univariant margins of
$\psi$ are given by {\it Gram matrices\,} formed by Hermitian dot
products of parallel slices of $[\psi_{ijk}]$.
It follows that $\rk\rho_C\le \rk\rho_A\cdot\rk\rho_B$, because
$\rho_C$ can be written as Gram matrix of the slices of dimension
$\rk\rho_A\cdot\rk\rho_B$. This inequality is a simplest
manifestation of general problem about relations between margins
of a pure state, which we address below.
\end{Ex}

\subsection{Marginal problem} \label{margin} {\it General quantum marginal problem\,} is
about existence of mixed state $\rho_I$ of composite system
$$\mathcal{H}_I=\bigotimes_{i\in I}\mathcal{H}_i$$
with given margins $\rho_J$ for some $J\subset I$ (cf. with
classical settings $n^\circ$\ref{class}).

Additional restrictions on state $\rho_I$ may be relevant. Here we
consider only two variations:
 \begin{itemize}
\item {\it Pure marginal problem\,}

\end{itemize}
 corresponding to pure state $\rho_I$, and more general
 \begin{itemize}
\item{\it Spectral marginal problem\,}
\end{itemize} corresponding to a state with
given spectrum $\lambda_I=\Spec\rho_I$.

Both versions are nontrivial even for univariant margins (cf. with
Gale--Ryser theorm $n^\circ$\ref{restr}). In this case margins
$\rho_i$ can be diagonalized by local unitary transformations
and their compatibility depends only on spectra
$\lambda_i=\Spec\rho_i$.

Pure quantum marginal problem has no classical analogue, since
projection of a point (=``pure state") is a point. In simplest
univariant case it can be stated in plain language as follows.
\begin{prob}\label{spectral} 
Let $M=[m_{ijk}]$ be complex cubic matrix and $M_1,M_2,M_3$ be
Gram matrices
 formed by Hermitian dot products of parallel slices
of $M$ (see Example \ref{slice}). What are possible spectra of
matrices $M_1$, $M_2$, and $M_3$?
\end{prob}


\subsection{Some known results}
In contrast with classical marginal problem its quantum
counterpart attracts attention only quite recently. Until  now
there were very few known results in this direction, which are
listed below.

\subsubsection{Pure two component problem} \label{pure}
 Equation \ref{2-comp}
implies that margins of pure state $\psi\in
\mathcal{H}_A\otimes\mathcal{H}_B$ are {\it isospectral\,}
$\Spec\rho_A=\Spec\rho_B$ if we discard zero eigenvalues. Vice
versa, this condition is sufficient for compatibility. Indeed let
$\alpha_i,\beta_i$ be orthonormal eigenbases of $\rho_A,\rho_B$
with common eigenvalues $\lambda_i$. Then state
$\psi=\sum_i\lambda_i\alpha_i\otimes\beta_i$ has margins
$\rho_A,\rho_B$. This representation of $\psi$ is known as {\it
Schmidt decomposition}.
\subsubsection{Scalar margins}\label{scalar}
Completely entangled state
$\psi\in\mathcal{H}_I=\bigotimes_i\mathcal{H}_i$ can be
characterized by its univariant margins $\rho_i$ \cite{VDMV,Kl02}:
$$\psi \mbox{ is completely entangled}\Longleftrightarrow
\rho_i=\mbox{scalar}.$$
\begin{thm*}[Klyachko \cite{Kl02}] Pure state $\psi\in\mathcal{H}_I$ with scalar univariant margins
exists iff informational capacities of the components
$\delta_i=\log\dim\mathcal{H}_i$ satisfy polygonal inequalities
$$\delta_i\le\sum_{j(\neq i)}\delta_j.$$
\end{thm*}

\subsubsection{Pure $N$-qubit problem} \label{N-qubit}
In this case there is a simple criterion for compatibility
univariant margins.
\begin{thm*}[Higuchi et al. \cite{HSS}, Bravyi \cite{Brav}]  Pure
$N$-qubit state $\psi\in\mathcal{H}^N,
\dim
\mathcal{H}=2$ with univariant margins $\rho_i$ exists iff their minimal
eigenvalues $\lambda_i$
satisfy polygonal inequalities
\begin{equation}\label{poly}\lambda_i\le\sum_{j(\neq i)}\lambda_j.
\end{equation}
\end{thm*}

\subsubsection{Mixed $2$-qubit problem}\label{two_qubit}
This problem was solved by Sergey Bravyi.
\begin{thm*}[Bravyi  \cite{Brav}]
Mixed two-qubit state $\rho_{AB}$ with spectrum
$\lambda_1\ge\lambda_2\ge\lambda_3\ge\lambda_4$ and margins
$\rho_A,\rho_B$ exists iff minimal eigenvalues
$\lambda_A,\lambda_B$ of the margins satisfy
inequalities
\begin{align}\lambda_A&\ge\lambda_3+\lambda_4,\quad\lambda_B\ge\lambda_3+\lambda_4,\nonumber\\
\lambda_A&+\lambda_B\ge\lambda_2+\lambda_3+2\lambda_4,\label{brav}\\
|\lambda_A&-\lambda_B|\le\min(\lambda_1-\lambda_3,\lambda_2-\lambda_4).\nonumber
\end{align}
\end{thm*}
\subsubsection{Pure 3-qutrit problem}\label{hig}
Astuchi Higuchi  found a criterion for compatibility of univarint
margins in 3-qutrit system
$\mathcal{H}_A\otimes\mathcal{H}_B\otimes\mathcal{H}_C$,
$\dim\mathcal{H}_*=3$ in terms of marginal spectra
$\lambda_1^*\le\lambda_2^*\le\lambda_3^*$, $*=A,B,C$.
\begin{thm*}[Higuchi \cite{Hig}] Pure state
$\psi\in\mathcal{H}_A\otimes\mathcal{H}_B\otimes\mathcal{H}_C$
with  margins $\rho_A,\rho_B,\rho_C$ exists iff the following
inequalities holds
\begin{align*}
\lambda_2^a+\lambda_1^a&\le\lambda_2^b+\lambda_1^b+
\lambda_2^c+\lambda_1^c,\\
\lambda_3^a+\lambda_1^a&\le\lambda_2^b+\lambda_1^b+
\lambda_3^c+\lambda_1^c,\\
\lambda_3^a+\lambda_2^a&\le\lambda_2^b+\lambda_1^b+
\lambda_3^c+\lambda_2^c,\\
2\lambda_2^a+\lambda_1^a&\le2\lambda_2^b+\lambda_1^b+
2\lambda_2^c+\lambda_1^c,\\
2\lambda_1^a+\lambda_2^a&\le2\lambda_2^b+\lambda_1^b+
2\lambda_1^c+\lambda_2^c,\\
2\lambda_2^a+\lambda_3^a&\le2\lambda_2^b+\lambda_1^b+
2\lambda_2^c+\lambda_3^c,\\
2\lambda_2^a+\lambda_3^a&\le2\lambda_1^b+\lambda_2^b+
2\lambda_3^c+\lambda_2^c,
\end{align*}
where $a,b,c$ is a permutation of $A,B,C$.
\end{thm*}
It takes 46 pages to prove this!
\subsubsection{Remark} All the above theorems deal with univariant margins. In
quantum field theory they are known as {\it mean fields\,}, and
higher rank margins as {\it $n$-point correlations\,}. Most
physical effects are governed by two-point correlations. However
complete solution of bivariant marginal problem is hardly possible
(even in classical case, see $n^\circ$\ref{Planar}). Here is a
couple of sporadic facts beyond trivial
compatibility relations like
$\Tr_A\rho_{AB}=\rho_B=\Tr_C\rho_{BC}$.

\subsubsection{}
Strong subadditivity \cite{LR,HJPW} of quantum entropy
$S(\rho)=-\Tr(\rho\log\rho)$
$$S(\rho_{ABC})\le S(\rho_{AB})+S(\rho_{BC})-S(\rho_B)$$
imposes a restriction on bivariant margins.

\subsubsection{} There is no pure 4-qubit state with scalar bivariant
margins \cite{HS}.

\section{Marginal inequalities}\label{mineq}
In this section we give a general recipe
for producing marginal inequalities for arbitrary multi-component
system based on Berenstin--Sjamaar theorem \cite{Ber-Sja}. For $n$
qubits our result, modulo a ``standard" conjecture, amounts to a
simple combinatorics, but the number of involved inequalities
increases drastically with $n$. The marginal inequalities up to 4
qubits, and for systems of formats $2\times3$, $3\times3$,
$2\times4$, and $2\times2\times3$ are given in Appendix.

\subsection{Main result}\label{MargIneq2}
To avoid technicalities we confine  ourselves to two component
system $\mathcal{H}_{AB}=\mathcal{H}_{A}\otimes\mathcal{H}_{B}$.
Generalization to multicomponent systems is straightforward. We
start with some auxiliary notions and results.
\subsubsection{Filtrations} Let $\alpha$ be a {\it nonincreasing filtration\,} of space
$\mathcal{H}$, i.e.  one parametric system of subspaces
$\mathcal{H}^\alpha(t)\subset\mathcal{H}, t\in\mathbb{R}$ such
that $\mathcal{H}^\alpha(s)\subset\mathcal{H}^\alpha(t)$  for
$s\ge t$ and
\begin{equation}\mathcal{H}^\alpha(t)=
\begin{cases}\mathcal{H},& \text{for $t\ll0$},\\
                      0,&\text{for $t\gg0$}.
\end{cases}
\end{equation}
The filtration supposed to be left continuous:
$\lim_{\varepsilon\rightarrow+0}\mathcal{H}^\alpha(t-\varepsilon)=\mathcal{H}^\alpha(t)$.

Denote by
\begin{equation}\mathcal{H}^{[\alpha]}(t)=\lim_{\varepsilon\rightarrow+0}
\frac{\mathcal{H}^\alpha(t)}{\mathcal{H}^\alpha(t+\varepsilon)}\end{equation}
{\it composition factors\,} of the filtration. The dimension
$m_\alpha(t)=\dim\mathcal{H}^{[\alpha]}(t)$ called {\it
multiplicity\,} of $t$ in spectrum of $\alpha$. Finite set of real
numbers $t\in\mathbb{R}$ with positive multiplicity
$m_\alpha(t)>0$ called  {\it spectrum\,} of filtration $\alpha$.
This are just discontinuity points of {\it dimension function}
\begin{equation}
d_\alpha(t)=\dim\mathcal{H}^\alpha(t).
\end{equation}
We always arrange the spectral values (counted according the
multiplicities) in nonincreasing order
\begin{equation}\label{spec}\Spec(\alpha):a_1\ge a_2\ge\cdots\ge a_n,\quad
n=\dim\mathcal{H}_A.\end{equation}
\begin{Ex} Hermitian operator
$\alpha:\mathcal{H}\rightarrow\mathcal{H}$ defines {\it spectral
filtration\,}
$$\mathcal{H}^\alpha(t)=\left\{\parbox{5cm}{subspace
spanned by eigenspaces of $\alpha$ with eigenvalues $\ge
t$}\right\}.$$ Spectrum of this filtration is equal to spectrum of
the operator $\alpha$. Every filtration  of Hilbert space
$\mathcal{H}$ is a spectral one. Hence filtration is a metric
independent substitution for Hermitian operator.
\end{Ex}
We often refer to spectrum of a filtration  as its {\it type}.
Filtrations $\alpha$ of fixed type $a=\Spec(\alpha)$ form {\it
flag variety\,} $\Flag_{a}(\mathcal{H})$
consisting of all chains of subspaces
\begin{equation}\label{flag}\mathcal{H}=F^1\supset
F^2\supset\cdots\supset F^\ell\supset F^{\ell+1}=0,\end{equation}
with composition factors $F^i/F^{i+1}$ of dimension equal to
multiplicity of $i$-th spectral value in $a$.

\subsubsection{Composition of filtrations}
Let now $\alpha$, $\beta$ be filtrations in spaces
$\mathcal{H}_A$, $\mathcal{H}_B$ respectively. Define their {\it
composition\,} $\alpha\beta$ as filtration of
$\mathcal{H}_{AB}=\mathcal{H}_A\otimes\mathcal{H}_B$ given by
equation
\begin{equation}\mathcal{H}_{AB}^{\alpha\beta}(t)=\sum_{r+s=t}
\mathcal{H}_A^\alpha(r)\otimes\mathcal{H}_B^\beta(s).\label{comp}\end{equation}
The type of the composition $\alpha\beta$ depends only on spectra
$a=\Spec(\alpha)$ and $b=\Spec(\beta)$
\begin{equation}\label{SpecComp}\Spec(\alpha\beta)=\{a_i+b_j
\text{ arranged in nonincreasing order}\}.
\end{equation}
Therefore the composition of filtrations of {\it fixed\,} types
$a$  and $b$ gives rise to morphism of flag varieties
\begin{equation}\label{morph}
\varphi_{a,b}:
\Flag_{a}(\mathcal{H}_A)\times\Flag_{b}(\mathcal{H}_B)\rightarrow
\Flag_{ab}(\mathcal{H}_{AB}),\quad
\alpha\times\beta\mapsto\alpha\beta,
\end{equation}
where composition of spectra $ab$ is defined by RHS of equation
(\ref{SpecComp}).

\subsubsection{Cubicles and extremal edges} Morphism (\ref{morph}) depends only on
the {\it order\,} of quantities $a_i+b_j$ in spectrum
(\ref{SpecComp}). We call pairs of spectra $(a,b)$ and
$(\tilde{a},\tilde{b})$ to be equivalent iff they define the same
morphism $\varphi_{a,b}=\varphi_{\tilde{a},\tilde{b}}$. This means
that the quantities $a_i+b_j$ and $\tilde{a}_i+\tilde{b}_j$ come
in the same order
\begin{equation}a_i+b_j\le a_k+b_l\Leftrightarrow
\tilde{a}_i+\tilde{b}_j\le\tilde{a}_k+\tilde{b}_l.
\end{equation}
Note that affine transformations of the spectra
\begin{equation}
a_i\mapsto p\cdot a_i+q,\qquad b_j\mapsto p\cdot b_j+r,\quad
p>0;q,r\in\Real,
\end{equation}
preserve the equivalence classes. This allows reduce the spectra
$a,b$ to Weyl chambers
\begin{equation}\label{Weyl}
\begin{aligned}
\Delta_m&:a_1\ge a_2\ge\cdots\ge a_m,&{\sum}_i a_i&=0,& m&=\dim \mathcal{H}_A;\\
\Delta_n&:b_1\ge b_2\ge\cdots\ge b_n,&{\sum}_j b_j&=0,&n&=\dim \mathcal{H}_B.
\end{aligned}
\end{equation}
The equivalence produces a decomposition of the product
$\Delta_m\times\Delta_n$ into relatively open polyhedral cones. We
are mostly interested in cones of maximal and minimal dimension,
called {\it cubicles\,} and {\it extremal edges\,} respectively.
The cubicles  are just pieces into which hyperplanes
\begin{equation}\label{facet}
a_i+b_j=a_k+b_l.
\end{equation}
cut $\Delta_m\times\Delta_n$. Extremal edges are given by a system
of equations of this form with one dimensional space of solutions.

A cubicle consists of spectra $(a,b)$ with pairwise distinct
quantities $a_i+b_j$ coming in {\it fixed  order\,}. It can be
described by $m\times n$ matrix $T$ with entries $1,2,\ldots,mn$
written in the {\it opposite\,} order to that of matrix
$[a_i+b_j]$:
$$T_{ij}\ge T_{kl}\Leftrightarrow a_i+b_j\le
a_k+b_l.$$ In other words, $T$
shows a way of counting entries of matrix $[a_i+b_j]$ in {\it
decreasing\,} order. The entries of $T$ strictly increase in rows
and columns, i.e. $T$ is a {\it standard tableau\,} of rectangular
shape $m\times n$.
There is a famous {\it hook formula\,} \cite{Macdonald} for the
number of such tableaux, which in
current settings takes form
\begin{equation}\label{hook}
\#\{\text{ standard $m\times n$
tableaux }\}=\frac{(mn)!}{\prod\limits_{\substack{1\le i\le
m\\1\le j\le n}}(i+j-1)}
\end{equation}
and gives an {\it upper bound\,} for the number of cubicles.
\begin{Ex} \label{2xn} For system of format $2\times n$ every standard tableau
corresponds to a cubicle. Hence by (\ref{hook}) the number of
cubicles is equal to {\it Catalan number\,}
\begin{equation*}\label{Catalan} \#\{\text{ cubicles
}\}=\frac{1}{n+1}{2n\choose n}.
\end{equation*}
A typical extremal edge in this case  comes from spectra
$$a=(1/2,-1/2),\qquad b=(b_1,b_2,\ldots,b_n),$$
with $b_i-b_{i+1}=0,1; \quad\sum b_i=0$.
The remaining extremal edges are that of $\Delta_n$
$$ a=(0,0),\qquad b=(\underbrace{k,k,\ldots,k}_{n-k},\underbrace{k-n,k-n,\ldots,k-n}_k).$$
This amounts altogether to $2^{n-1}+n-1$ extremal edges, which is
of order square root of the number of cubicles. In simplest case
of two qubits they are
\begin{equation}\label{2x2}
\begin{aligned}
a&=(1,-1);&a&=(0,\;\;\,0);&a&=(1,-1);\\
b&=(1,-1);&b&=(1,-1);&b&=(0,\;\;\, 0);
\end{aligned}
\end{equation}
\end{Ex}
\begin{Ex}
For two qutrits (=system of format $3\times3$) there are $36$
cubicles represented by the following tableaux and their transpose
\begin{center}
\vspace{3mm}
\begin{tabular}{|c|c|c|}
\hline
1&4&7\\
\hline
2&5&8\\
\hline
3&6&9\\
\hline
\end{tabular}
\hfill
\begin{tabular}{|c|c|c|}
\hline
1&4&6\\
\hline
2&5&8\\
\hline
3&7&9\\
\hline
\end{tabular}
\hfill
\begin{tabular}{|c|c|c|}
\hline
1&4&6\\
\hline
2&5&7\\
\hline
3&8&9\\
\hline
\end{tabular}
\hfill
\begin{tabular}{|c|c|c|}
\hline
1&4&5\\
\hline
2&6&8\\
\hline
3&7&9\\
\hline
\end{tabular}
\hfill
\begin{tabular}{|c|c|c|}
\hline
1&4&5\\
\hline
2&6&7\\
\hline
3&8&9\\
\hline
\end{tabular}
\hfill
\begin{tabular}{|c|c|c|}
\hline
1&5&6\\
\hline
2&3&8\\
\hline
4&7&9\\
\hline
\end{tabular}

\vspace{3mm}
\begin{tabular}{|c|c|c|}
\hline
1& 3&6\\
\hline
2&5&7\\
\hline
4&8&9\\
\hline
\end{tabular}
\hfill
\begin{tabular}{|c|c|c|}
\hline
1&3&7\\
\hline
2&5&8\\
\hline
4&6&9\\
\hline
\end{tabular}
\hfill
\begin{tabular}{|c|c|c|}
\hline
1&3&7\\
\hline
2&4&8\\
\hline
5&6&9\\
\hline
\end{tabular}
\hfill
\begin{tabular}{|c|c|c|}
\hline
1&3&6\\
\hline
2&4&8\\
\hline
5&7&9\\
\hline
\end{tabular}
\hfill
\begin{tabular}{|c|c|c|}
\hline
1&3&6\\
\hline
2&4&7\\
\hline
5&8&9\\
\hline
\end{tabular}
\hfill
\begin{tabular}{|c|c|c|}
\hline
1&2&6\\
\hline
3&4&8\\
\hline
5&7&9\\
\hline
\end{tabular}

\vspace{3mm}
\begin{tabular}{|c|c|c|}
\hline
1&2&7\\
\hline
3&4&8\\
\hline
5&6&9\\
\hline
\end{tabular}
\hfill
\begin{tabular}{|c|c|c|}
\hline
1&2&5\\
\hline
3&6&8\\
\hline
4&7&9\\
\hline
\end{tabular}
\hfill
\begin{tabular}{|c|c|c|}
\hline
1&2&6\\
\hline
3&5&8\\
\hline
4&7&9\\
\hline
\end{tabular}
\hfill
\begin{tabular}{|c|c|c|}
\hline
1&2&7\\
\hline
3&5&8\\
\hline
4&6&9\\
\hline
\end{tabular}
\hfill
\begin{tabular}{|c|c|c|}
\hline
1&3&5\\
\hline
2&6&7\\
\hline
4&8&9\\
\hline
\end{tabular}
\hfill
\begin{tabular}{|c|c|c|}
\hline
1&3&5\\
\hline
2&6&8\\
\hline
4&7&9\\
\hline
\end{tabular}
\end{center}
\vspace{3mm}
By hook formula (\ref{hook}) there are 42 standard $3\times3$
tableaux, 6 of them correspond to no cubicle.
There are 17
extremal edges spanned by spectra $(a,b)$ or $(b,a)$ below
\begin{align*}
a&=(1,\;\;\,0,-1);&a&=(3,\;\;\,0,-3);&a&=(0,\;\;\,0,\;\;\,0);&a&=(3,\;\;\,0,-3);\\
b&=(1,\;\;\,0,-1);&b&=(5,-1,-4);&b&=(2,-1,-1);&b&=(1,\;\;\,1,-2);\\[3mm]
a&=(2,-1,-1);&a&=(3,\;\;\,0,-3);&a&=(0,\;\;\,0,\;\;\,0);&a&=(3,\;\;\,0,-3);\\
b&=(2,-1,-1);&b&=(4,\;\;\,1,-5);&b&=(1,\;\;\,1,-2);&b&=(2,-1,-1);\\[3mm]
&&a&=(1,\;\;\,1,-2);&a&=(2,-1,-1);&&\\
&&b&=(1,\;\;\,1,-2);&b&=(1,\;\;\,1,-2).&&
\end{align*}
\end{Ex}


\subsubsection{Cohomology of flag varieties}\label{cohom}
Let $T$ be a cubicle. By definition, morphism $\varphi_{ab}$
defined by equation (\ref{morph}) is independent of $(a,b)\in T$.
Hence
the notation $\varphi_{ab}=\varphi_T$.
Besides, all three spectra $a,b,ab$ are simple and the
corresponding flag varieties $\Flag_a(\mathcal{H}_A)$,
$\Flag_b(\mathcal{H}_B)$, $\Flag_{ab}(\mathcal{H}_{AB})$ are {\it
complete\,}, i.e. consist of filtrations with composition factors
of dimension one. Thus every cubicle gives rise to a well defined
morphism of complete flag varieties
\begin{equation}
\varphi_T:
\Flag(\mathcal{H}_A)\times\Flag(\mathcal{H}_B)\rightarrow
\Flag(\mathcal{H}_{AB})
\end{equation}
 and that of their cohomology
\begin{equation}\label{CohomMorph}
\varphi_T^*:
H^*(\Flag(\mathcal{H}_{AB}))\rightarrow
H^*(\Flag(\mathcal{H}_A))\otimes H^*(\Flag(\mathcal{H}_B)).
\end{equation}
Cohomology ring of complete flag variety $H^*(\Flag(\mathcal{H}))$
is generated by characteristic classes $x_i=c_1(\mathcal{L}_i)$ of
line bundles $\mathcal{L}_i$ with fibers equal to $i$-th
composition factor of the flag (\ref{flag}). We call $x_i$ {\it
canonical generators\,}. Elementary symmetric functions
$\sigma_i(x)$ of the canonical generators are characteristic
classes of trivial bundle $\mathcal{H}$ and thus vanish. This
identifies the cohomology ring with factor
\begin{equation}\label{canonical}
H^*(\Flag(\mathcal{H}))=\mathbb{Z}[x_1,x_2,\ldots,x_n]/(\sigma_1,\sigma_2,\ldots,\sigma_n).
\end{equation}
In term of the canonical generators $x_i,y_j,z_k$  of cohomology
of flag varieties $\Flag(\mathcal{H}_A)$, $\Flag(\mathcal{H}_B)$,
and $\Flag(\mathcal{H}_{AB})$ morphism $\varphi_T^*$ can be
described as follows
\begin{equation}\label{special}
\varphi_T^*: z_k\mapsto x_i+ y_j \text{ for } k=T_{ij},
\end{equation}
where we identify cubicle with the corresponding tableau $T$ and
for simplicity write $x_i,y_j$ instead of $x_i\otimes 1, 1\otimes
y_j$. In other words, $z_k\mapsto x_i+y_j$ iff $k$-th term of the
composition $ab$ is $a_i+b_j$ for any $(a,b)$ in cubicle $T$.

The cohomology ring $H^*(\Flag(\mathcal{H}))$ has a natural
geometric basis consisting of so called {\it Schubert cocycles\,}
$\sigma_w$, where $w\in S_n$ is a permutation of degree $n=\dim
\mathcal{H}$. They can be expressed via
characteristic classes $x_i$ in terms of {\it difference
operators\,}
\begin{equation}
\partial_i:f(x_1,x_2,\ldots,x_n)\mapsto
\frac{f(\ldots,x_i,x_{i+1},\ldots)-f(\ldots,x_{i+1},x_i,\ldots)}
{x_i-x_{i+1}}
\end{equation}
as follows. Write permutation $w\in S_n$ as product of minimal
number of transpositions $s_i=(i,i+1)$
\begin{equation}\label{decomp}
w=s_{i_1}s_{i_2}\cdots s_{i_\ell}.
\end{equation}
The number of factors $\ell(w)=\#\{i<j\mid w(i)>w(j)\}$ called
{\it length\,} of permutation $w$.
The product
\begin{equation}
\partial_w:=\partial_{i_1}\partial_{i_2}\cdots\partial_{i_\ell}
\end{equation}
is independent of reduced decomposition (\ref{decomp}) and  in
terms of these operators Schubert cocycle $\sigma_w$ is given by
equation
\begin{equation}\label{Schub}
\sigma_w=\partial_{w^{-1}w_0}(x_1^{n-1}x_2^{n-2}\cdots x_{n-1}),
\end{equation}
where  $w_0=(n,n-1,\ldots,2,1)$ is unique permutation of maximal
length.

Right hand side of equation (\ref{Schub}) called {\it Schubert
polynomial\,} $S_w(x_1,x_2,\ldots,x_n)$, $\deg S_w=\ell(w)$. These
polynomials where first introduced by Lascoux and
Sch\"utzen\-berger \cite{L-Sch} who studied them in a long series
of papers. See \cite{Mac91} for further references and a concise
exposition of the theory. We borrowed  from \cite{L-Sch}  the
following table, in which $x,y,z$ stand for $x_1,x_2,x_3$. 
\begin{center}{\small\label{SchPol}
\begin{tabular}[b]{|c|c||c|c||c|c||c|c|}
\hline
$w$ & $S_w$&$w$&$S_w$&$w$&$S_w$&$w$&$S_w$\\
\hline%
3201&{\footnotesize $x^3y^2z$}&2301&{\footnotesize $x^2y^2$}&
2031&{\footnotesize $x^2y+x^2z$}&1203&{\footnotesize $xy$}\\
\hline
2310&{\footnotesize $x^2y^2z$}&3021&{\footnotesize $x^3y+x^3z$}&
2103&{\footnotesize $x^2y$}&2013&{\footnotesize $x^2$}\\
\hline
3120&{\footnotesize $x^3yz$}&3102&{\footnotesize $x^3y$}&
3012&{\footnotesize $x^3$}&0132&{\footnotesize $x+y+z$}\\
\hline
3201&{\footnotesize $x^3y^2$}&1230&{\footnotesize $xyz$}&
0231&{\footnotesize $xy+yz+zx$}&0213&{\footnotesize $x+y$}\\
\hline
1320&{\footnotesize $x^2yz+xy^2z$}&0321&{\footnotesize $x^2y+x^2z+xy^2$}&
0312&{\footnotesize $x^2+xy+y^2$}&1023&{\footnotesize $x$}\\
\hline
2130&{\footnotesize $x^2yz$}&1302&{\footnotesize $x^2y+xy^2$}&
1032&{\footnotesize $x^2+xy+xz$}&0123&{\footnotesize $1$}\\
\hline
\end{tabular}}
\end{center}

Extra variables $x_{n+1},x_{n+2},\ldots$ being added to
(\ref{Schub}) leave Schubert polynomials unaltered.
By that reason they are usually treated as
polynomials in infinite alphabet  $X=(x_1,x_2,\ldots)$. With this
understanding every homogeneous polynomial can be decomposed into
Schubert components
as follows
\begin{equation}
f(X)=\sum_{\ell(w)=\deg(f)} \partial_w f\cdot S_w(X).
\end{equation}
Applying this to specialization (\ref{special}) of Schubert
polynomial $S_w(Z)$ we get decomposition
\begin{equation}\label{SchMap}
\varphi_T(S_w(Z))=\sum_{\ell(u)+\ell(v)=\ell(w)} c_{uv}^w(T)\cdot
S_u(X)S_v(Y),
\end{equation}
where
\begin{equation}\label{SchCoeff}
c_{uv}^w(T)=\partial_u\partial_v
S_w(Z)\left|_{z_k=x_i+y_j}\right.,\qquad k=T_{ij}.
\end{equation}
Here operators $\partial_u,\partial_v$ act on variables $x,y$
respectively. Reduction of (\ref{SchMap}) modulo elementary
symmetric functions in $x$ and  in $y$ gives cohomology morphism
(\ref{CohomMorph}) in terms of Schubert cocycles
\begin{equation}\label{CocycleMap}
\varphi_T(\sigma_w)=\sum_{\ell(u)+\ell(v)=\ell(w)} c_{uv}^w(T)\cdot
\sigma_u\otimes\sigma_v.
\end{equation}
\subsubsection{Marginal inequalities} Now we are in position to
state a solution of the quantum marginal problem.
\begin{thm}\label{cubicleThm}
Let $\mathcal{H}_{AB}=\mathcal{H}_A\otimes \mathcal{H}_B$ be two
component system.
Then   the following
conditions on spectra $\lambda^A,\lambda^B,\lambda^{AB}$
are equivalent.
\newline
${}$\qquad {\rm (1)} There exists mixed state $\rho_{AB}$ with
margins $\rho_A$, $\rho_B$ and spectra
$\lambda^{AB},\lambda^A,\lambda^B$.
\newline
${}$\qquad {\rm (2)}
Each time the coefficient $c_{uv}^w(T)$ of decomposition
$(\ref{CocycleMap})$
is nonzero, the following inequality holds for all $(a,b)$ in the
cubicle $T$
\begin{equation}\label{MargIneq}
\sum_i a_i\lambda^A_{u(i)}+\sum_j b_j\lambda^B_{v(j)}\ge \sum_{k}
(a_i+b_j)_k\lambda^{AB}_{w(k)},
\end{equation}
where $(a_i+b_j)_k$ is $k$-th element of the sequence $a_i+b_j$
arranged in decreasing order.
\end{thm}
\begin{proof}
The quantum marginal problem amounts to decomposition of
projection of a coadjoint orbit of group
$\SU(\mathcal{H}_A\otimes\mathcal{H}_B)$ into coadjoint orbits of
subgroup $\SU(\mathcal{H}_A)\times\SU(\mathcal{H}_B)$.

The above discussion just puts the result into
framework of Berenstein-Sjamaar theorem \cite[Thm 3.2.1]{Ber-Sja}
applied  these groups.
\end{proof}
\begin{rem}
Every cubicle is interior of convex hull  of its extremal edges.
Therefore inequality (\ref{MargIneq}) enough to check for extremal
edges only.
\end{rem}
\begin{Ex} Note that $c_{uv}^w(T)=1$ for identical permutations $u,v,w$.
Hence the following inequality holds for every extremal edge
$(a,b)$
\begin{equation}\label{basic}
\sum_i a_i\lambda^A_{i}+\sum_j b_j\lambda^B_{j}\ge \sum_{k}
(a_i+b_j)_k\lambda^{AB}_{k}.
\end{equation}
We call these marginal inequalities {\it basic\,} ones.
\end{Ex}
\begin{Ex}\label{basic2x2} {\it ``Easy" two qubit inequalities.} Extremal edges for two qubits from Example \ref{2xn}
give rise to basic inequalities
\begin{align*}
&\lambda^A_1-\lambda^A_2+\lambda^B_1-\lambda^B_2\le 2(\lambda^{AB}_1-\lambda^{AB}_4),\\
&\lambda^A_1-\lambda^A_2\le \lambda^{AB}_1+\lambda^{AB}_2-\lambda^{AB}_3-\lambda^{AB}_4,\\
&\lambda^B_1-\lambda^B_2\le
\lambda^{AB}_1+\lambda^{AB}_2-\lambda^{AB}_3-\lambda^{AB}_4,
\end{align*}
which are equivalent to the first three inequalities in
(\ref{brav}). S.~Bravyi \cite{Brav} calls  them {\it easy\,} ones.
\end{Ex}
Extremal edge $(a,b)$ gives rise to
cohomology map
\begin{equation}\label{IncplMorph}
\varphi_{ab}^*:H^*(\Flag_{ab}(\mathcal{H}_{AB}))\rightarrow
H^*(\Flag_{a}(\mathcal{H}_{A}))\otimes
H^*(\Flag_{b}(\mathcal{H}_{B}))
\end{equation}
induced by morphism $\varphi_{ab}$
from equation (\ref{morph}). In contrast with cubicle case,
spectrum $ab$ is never simple and we have to deal with varieties
of {\it incomplete\,} flags.

Let $\alpha$ be a filtration of space $\mathcal{H}$ of dimension
$m$ and spectrum $a=\Spec\alpha$. Denote by
$\mu(a)=(m_1,m_2,\ldots,m_\ell)$, $\sum_i m_i=m$ the
multiplicities of the spectrum.
%
Recall that flag variety $\Flag_{a}(\mathcal{H})$ consists of
the filtrations with composition factors of dimension $m_i$. Its
cohomology
known to be subring of invarinats
\begin{equation}\label{IncFlag}
H^*(\Flag_{a}(\mathcal{H}))=H^*(\Flag(\mathcal{H}))^{S_\mu}\subset
H^*(\Flag(\mathcal{H}))
\end{equation}
with respect to Schur subgroup $S_\mu=S_{m_1}\times
S_{m_2}\times\cdots\times S_{m_\ell}\subset S_m$ acting on
cohomolohy of {\it complete\,} flags by permutations of the
canonical generators $x_i$, see \cite{BGG}. An important example
of such invariant comes from Schubert cocycle $\sigma_w$
corresponding to a {\it shuffle of type $\mu(a)$}, i.e.
permutation $w\in S_m$ which preserves order of the first $m_1$
elements, the next $m_2$ elements, and so on. Such Schubert
cocycles form a basis of the cohomology ring (\ref{IncFlag}) of
incomplete flags.

In view of this interpretation morphism (\ref{IncplMorph}) becomes
just a restriction of
\begin{equation}
\varphi_T^*:
H^*(\Flag(\mathcal{H}_{AB}))\rightarrow
H^*(\Flag(\mathcal{H}_A))\otimes H^*(\Flag(\mathcal{H}_B))
\end{equation}
onto the algebras of invariants for any cubicle $T$ containing
extremal edge $(a,b)$ in its closure. Hence by
(\ref{CocycleMap})-(\ref{SchCoeff})
coefficients of the decomposition
\begin{equation}\label{EdgeMorph}
\varphi_{ab}(\sigma_w)=\sum_{\ell(u)+\ell(v)=\ell(w)} c_{uv}^w(a,b)\cdot
\sigma_u\otimes \sigma_v,
\end{equation}
are given by equation
\begin{equation}
c_{uv}^w(a,b)=\partial_u\partial_v
S_w(Z)\left|_{z_k=x_i+y_j}\right.,\qquad k=T_{ij},
\end{equation}
where $u,v,w$ are shuffles of types $\mu(a),\mu(b),\mu(ab)$
respectively. This leads to the following description of marginal
inequalities in terms of extremal edges only.


\begin{thm}\label{ExtrEdge} All marginal inequalities for system
$\mathcal{H}_{AB}=\mathcal{H}_A\otimes \mathcal{H}_B$ can be
obtained from basic ones $(\ref{basic})$ by shuffles $u,v,w$ of
types $\mu(a),\mu(b),\mu(ab)$
\begin{equation}\label{modifiedINQ}
\sum_i a_i\lambda^A_{u(i)}+\sum_j b_j\lambda^B_{v(j)}\le\sum_{k}
(a_i+b_j)_k\lambda^{AB}_{w(k)}
\end{equation}
such that coefficient $c_{uv}^w(a,b)$ of decomposition
$(\ref{EdgeMorph})$ is nonzero.
\end{thm}
Extensive tables of marginal inequalities deduced from this
theorem are given in Addendum. Details of the calculation will be
published elsewhere \cite{Kl-Oz}.

\begin{Ex} {\it ``Difficult" two qubit inequality\,.}
Let's find modifications (\ref{modifiedINQ}) of the first basic
inequality form Example \ref{basic2x2}
\begin{equation}\label{first}
\lambda^A_1-\lambda^A_2+\lambda^B_1-\lambda^B_2\le
2(\lambda^{AB}_1-\lambda^{AB}_4),
\end{equation}
coming from extremal edge $a=(1,-1),b=(1,-1)$. There are two ways
of counting entries of matrix
\begin{equation*}\label{ab}
[a_i+b_j]=\begin{tabular}{|c|c|}
\hline
2&0\\
\hline
0&-2\\
\hline
\end{tabular}
\end{equation*}
in decreasing order
\begin{equation}\label{2x2cubicle}
\begin{tabular}{|c|c|}
\hline
1&3\\
\hline
2&4\\
\hline
\end{tabular}
\qquad \text{ or }\qquad
\begin{tabular}{|c|c|}
\hline
1&2\\
\hline
3&4\\
\hline
\end{tabular}
\end{equation}
which exhibit  two cubicles containing $(a,b)$ in theirs closure.

By Theorem \ref{ExtrEdge}
the modifications come from components of
$\varphi_{ab}^*(\sigma_w)$ for shuffles $w$ of type $(1,2,1)$ and
length $\ell(w)\le 2$. Using first cubicle in (\ref{2x2cubicle})
the calculation amounts to specialization (\ref{special})
$$z_1=x_1+y_1,\quad z_2=x_2+y_1, \quad z_3=x_1+y_2, \quad z_4=x_2+y_2  $$
of Schubert polynomial $S_w(Z)$ taken modulo relations
$$x_1+x_2=y_1+y_2=0,\qquad x_1x_2=y_1y_2=0.$$
From the table
on page \pageref{SchPol} it follows
\begin{gather*}
\varphi_{ab}^*(\sigma_{2134})=\varphi_{ab}^*(\sigma_{1243})=x_1+y_1=
\sigma_{21}\otimes\sigma_{12}+\sigma_{12}\otimes\sigma_{21},\\
\varphi_{ab}^*(\sigma_{2143})=\varphi_{ab}^*(\sigma_{3124})
=\varphi_{ab}^*(\sigma_{1342})=2x_1y_1=2\sigma_{21}\otimes\sigma_{21}.
\end{gather*}
The first line gives rise to 4 modified inequalities
\begin{align*}
\lambda^A_2-\lambda^A_1+\lambda^B_1-\lambda^B_2&\le
2(\lambda^{AB}_2-\lambda^{AB}_4),\\
\lambda^A_1-\lambda^A_2+\lambda^B_2-\lambda^B_1&\le
2(\lambda^{AB}_2-\lambda^{AB}_4),\\
\lambda^A_2-\lambda^A_1+\lambda^B_1-\lambda^B_2&\le
2(\lambda^{AB}_1-\lambda^{AB}_3),\\
\lambda^A_1-\lambda^A_2+\lambda^B_2-\lambda^B_1&\le
2(\lambda^{AB}_1-\lambda^{AB}_3),
\end{align*}
which can be squeezed into the following one
$${\mid(\lambda^A_1-\lambda^A_2)-(\lambda^B_1-\lambda^B_2)\mid}\le
2\min(\lambda^{AB}_1-\lambda^{AB}_3,\lambda^{AB}_2-\lambda^{AB}_4).$$
This is the last inequality in (\ref{brav})
designated by S.~Bravyi \cite{Brav} as a {\it difficult\,} one.
All the other modified inequalities are redundant.
We'll discuss the underlying reason for this in the next section.
\end{Ex}
\subsubsection{Redundancy}
Redundancy of marginal inequalities
is an important issue. For example, for two qutrits there are 17
extremal edges $(a,b)$, 2298 permutations $w\in S_9$ of length
$\le 12$, in addition $\varphi_{ab}^*(\sigma_w)$ may have many
components, and each gives a marginal inequality. Among those tens
of thousands inequalities only 397 are independent.
\begin{conj}\label{conj} All marginal constraints are given by
inequalities $(\ref{modifiedINQ})$ of theorem \ref{ExtrEdge} with
$c_{uv}^w=1$.
\end{conj}
This ``standard" conjecture may be true in framework of general
Berenstein-Sjamaar theorem \cite{Ber-Sja}. It is valid for
multicomponent quantum systems of rank $\le 4$, see Appendix.
Apparently the conjecture should follow from rigidity argument of
Belkale \cite{Bel}, who proved it for Hermitian spectral problem,
see section \ref{AddSpec}. In this case the inequalities listed in
the conjecture are also independent \cite{KTW}. However for
quantum marginal problem this is not the case. For example, for
system of format $2\times2\times3$ even some basic inequalities
are redundant. Redundancy also occurs in a counterpart of the
Hermitian spectral problem for groups other then $\SU(n)$. In this
case Belkale and Kumar \cite{Bel-Kum} have a refined procedure
which gives independent inequalities, at least in some examples.

\subsection{Arrays of qubits}\label{n_qubit} In this section we study in details
the system of $n$ qubit, which is of fundamental importance for
quantum information. In this case, modulo Conjecture \ref{conj},
the marginal inequalities can be written down  in a
straightforward manner from the list of extremal edges.
\subsubsection{Cubicles and extremal edges} Let's consider
array of qubits
$$\mathcal{H}^{\otimes n} = \mathcal{H}_1\otimes \mathcal{H}_2\otimes\cdots
\otimes \mathcal{H}_n,\quad\dim \mathcal{H}_i=2$$
and denote by  $\alpha_i$ a filtration of $\mathcal{H}_i$,
normalized to Weyl chamber (\ref{Weyl}), so that
$$\Spec\alpha_i=\pm a_i,\quad a_i\ge0.$$
Henceforth we'll identify the spectrum with nonnegative number
$a_i\ge0$. As in (\ref{morph}) composition of filtrations gives
rise to morphism
\begin{equation}\label{comp_morph}
\varphi_a:\Flag_{a_1}(\mathcal{H}_1)\times\Flag_{a_2}(\mathcal{H}_2)\times
\cdots\times\Flag_{a_n}(\mathcal{H}_n)\rightarrow
\Flag_{A}(\mathcal{H}_1\otimes\mathcal{H}_2\otimes\cdots\otimes\mathcal{H}_n),
\end{equation}
where on the left hand side $a$
stands for vector of spectra $(a_1,a_2,\ldots,a_n)$, while on the
right hand side $A$ refers to their composition
$$A=a_1a_2\cdots a_n=\left\{\pm a_1\pm a_2\pm\cdots\pm a_n
\text{ arranged in nonincreasing order}\right\}.$$
 Note that flag variety $\Flag_{a_i} (\mathcal{H}_i)$
amounts to projective line $\mathbb{P}^1$ (=Riemann sphere) for
$a_i>0$, or to a point for $a_i=0$. Similar to two component
system, $\varphi_a$ depends only on the order of $2^n$ quantities
\begin{equation}\label{proj}
\pm a_1\pm a_2\pm \cdots\pm a_n
\end{equation}
which are just dot products
of vector $a$ and a vertex $(\pm1,\pm1,\ldots,\pm1)$ of $n$-cube.
Hence $n$-qubit  cubicle is determined by the order of pairwise
distinct projections of vertices of the $n$-cube onto positive
direction $a$. Equivalently, cubicles are pieces into which
hyperplanes
\begin{equation}\label{wall}
\varepsilon_1a_1+\varepsilon_1a_1+\cdots+\varepsilon_n a_n=
\delta_1a_1+\delta_1a_1+\cdots+\delta_n a_n,\quad\varepsilon,\delta=\pm1,
\end{equation}
cut positive octant. Equation of the wall (\ref{wall}) separating
two cubicles can be recast in the form
\begin{equation}\label{wall2}
H_\eta:\eta_1a_1+\eta_2a_2+\cdots+\eta_na_n=0,\quad\eta_i=0,\pm1,
\end{equation}
which tells that  $H_\eta$ is orthogonal to vector $\eta$ pointing
to center of a face of the cube.

Every extremal edge is intersection of $n-1$ walls, hence given by
an independent system of homogeneous equations with $(n-1)\times
n$ matrix $M$ filled by $0,\pm 1$. The solution is given by a
vector of maximal
minors taken with alternating signs $(-1)^iM_i$. Their absolute
values
\begin{equation*}
(\abs{M_1},\abs{M_2},\ldots,\abs{M_n})
\end{equation*}
gives a positive solution for another such matrix $M^\prime$ and
spans an extremal edge.
 Computer implementation of this brute
force approach allows to find all extremal edges up to 6 qubits.
The table below shows the number of extremal edges counted up to a
permutation of qubits, say normalized by condition $a_1\le a_2\le
\ldots\le a_n$.

\begin{center}
\vspace{1mm}
\begin{tabular}{|c|c|c|c|c|c|}
\hline
\# qubits &2&3&4&5&6\\
\hline
\# edges&2&4&12&125&11344\\
\hline
\end{tabular}
\end{center}

\begin{Ex} {\it Three qubits.}  Let $0\le a\le b \le c$ be three spectra
in increasing order. Then among 8 quantities $\pm a\pm b
\pm c$ only one inequality is indefinite
\begin{equation*}
a+b+c\ge-a+b+c\ge a-b+c\ge a+b-c\gtrless-a-b+c\ge -a+b-c\ge
a-b-c\ge -a-b-c
\end{equation*}
and depends on whether $a,b,c$ satisfy triangle inequality or not.
In the former case the order is
$$a+b+c\ge-a+b+c\ge a-b+c\ge
a+b-c\ge-a-b+c\ge -a+b-c\ge a-b-c\ge -a-b-c,$$ and in the later
$$a+b+c\ge-a+b+c\ge a-b+c\ge-a-b+c \ge a+b-c\ge -a+b-c\ge
a-b-c\ge -a-b-c.$$ Thus the system has two cubicles up to
permutations of qubits and 12 altogether. The two cubicles
have 4 extremal edges
\begin{equation}\label{Edges3qbt}
(0,0,1);\quad(0,1,1);\quad(1,1,1);\quad (1,1,2).
\end{equation}
\end{Ex}
\begin{Ex} The 12 extremal edges $a_1\le a_2\le a_3\le a_4$ for 4
qubits are as follows
\begin{align*}
&(0,0,0,1),&&(0,0,1,1),&&(0,1,1,1),&&(0,1,1,2),\\
&(1,1,1,1),&&(1,1,1,2),&&(1,1,1,3),&&(1,1,2,2),\\
&(1,1,2,3),&&(1,1,2,4),&&(1,2,2,3),&&(1,2,3,4).
\end{align*}
\end{Ex}

\begin{rem}
It would be nice to understand better geometry and combinatorics
of the system of hyperplanes $H_\eta$.
Note that in a configuration of $N$ central hyperplanes in general
position there are ${N\choose n-1}$ lines of intersections. This
gives an {\it upper bound\,} for the number of lines of
intersection in the system of $(3^n-1)/2$ walls (\ref{wall2}) and
leads to estimate
\begin{equation*}
\sum_{\substack{\text{extremal edges}\\a_1\le a_2\le\cdots\le a_n}}
\frac{1}{\mid\text{Stab}\,(a)\mid}
\le \frac{1}{2^{n-1}n!}{(3^n-1)/2\choose n-1},
\end{equation*}
where $\text{Stab}\,(a)$ is stabilizer of edge $a$ in group of
monomial permutations $a_i\mapsto \pm a_j$. In worst case scenario
it may happens that the system $H_\eta$ becomes stochastic
for $n\gg 1$. Then there may be no reasonable description of its
edges and cubicles, except statistical one.

\end{rem}
\subsubsection{Marginal inequalities} Every vector $a$ in cubicle $T$
gives rise to morphism (\ref{comp_morph}) of {\it complete\,} flag
varieties depending only on $T$
\begin{equation}
\varphi_T:\mathbb{P}(\mathcal{H}_1)\times
\mathbb{P}(\mathcal{H}_2)\times\cdots\times\mathbb{P}(\mathcal{H}_n)\rightarrow
\Flag(\mathcal{H}_1\otimes\mathcal{H}_2\otimes\cdots\otimes\mathcal{H}_n),
\end{equation}
where we identify flag variety of $i$-th qubit
with projective line $\mathbb{P}(\mathcal{H}_i)$. By theorem
\ref{cubicleThm} marginal inequalities come from the induced
cohomology morphism
\begin{equation}
\varphi_T^*:H^*(\Flag(\mathcal{H}_1\otimes\mathcal{H}_2\otimes\cdots\otimes\mathcal{H}_n))
\rightarrow \mathbb{Z}[x_1,x_2,\ldots,x_n]/(x_i^2=0).
\end{equation}
Here we use isomorphism
$H^*(\mathbb{P}(\mathcal{H}_i))\cong\mathbb{Z}[x_i]/(x_i^2=0)$,
where $x_i=\sigma_{21}$ is Schubert cocycle.
In terms of canonical generators (\ref{canonical}) the morphism
amounts to specialization
\begin{equation}\label{SpecQbt}
\varphi_T^*:z_k\mapsto \pm x_1\pm x_2\pm\cdots \pm x_n,
\end{equation}
where the signs are taken from $k$-th term of the sequence $\pm
a_1\pm a_2\pm\cdots \pm a_n$ arranged  in decreasing order.
Applying this to Schubert cocycle $\sigma_w$ we get decomposition
\begin{equation}
\varphi_T^*(\sigma_w)=\sum c^w_{b_1b_2\ldots
b_n}(T)\cdot x_1^{b_1}x_2^{b_2}\ldots x_n^{b_n}
\end{equation}
into sum of square free monomials with exponents ${b_i=0,1}$. The
theorem \ref{cubicleThm} tells that each time $c^w_{b_1b_2\ldots
b_n}(T)$ is nonzero the following inequality holds for all $a\in
T$
\begin{equation}\label{QubitIneq}
\sum_i (-1)^{b_i}a_i(\rho_1^{(i)}-\rho_2^{(i)})\le \sum_{k}(\pm
a_1\pm a_2\pm\cdots\pm a_n)_k \rho_{w(k)},
\end{equation}
where $\rho$ is  $n$ qubit state with spectrum
$\rho_1\ge\rho_2\ge\cdots\ge \rho_{2^n} $ and margins
$\rho^{(i)}$, $(\pm a_1\pm a_2\pm\cdots\pm a_n)_k$ is $k$-th term
of the sequence $\pm a_1\pm a_2\pm\cdots\pm a_n$ arranged in
decreasing order, and sign $(-1)^{b_i}$ reflects the effect of
transposition of eigenvalues $\rho_1^{(i)}\ge
\rho_2^{(i)}$  of reduced state $\rho^{(i)}$.

Recall that according to Conjecture \ref{conj} all marginal
constraints come from inequalities (\ref{QubitIneq}) with
$c^w_{b_1b_2\ldots b_n}(T)=1$.
\begin{claim}\label{claim} For $n$ qubit system components of multiplicity one
in $\varphi_T^*(\sigma_w)$ can appear only for identical
substitution $w=1$ or for odd transposition $w=(2j-1,2j)$.
\end{claim}
\begin{proof} Indeed $\varphi_T^*(\sigma_w)$ is obtained from
Schubert polynomial $S_w(Z)$ by specialization (\ref{SpecQbt}).
Since $\pm x_1\pm x_2\pm \cdots\pm x_n\equiv x_1+x_2+\cdots+ x_n
\mod 2$, then
\begin{equation}\label{mod2}
\varphi^*_T(\sigma_w)\equiv
S_w(1,1,\ldots,1)(x_1+x_2+\cdots+x_n)^{\ell(w)}\mod 2.
\end{equation}
Let $\ell(w)=\sum_\alpha 2^\alpha$ be binary decomposition of
$\ell(w)$. Then
\begin{equation}
(x_1+x_2+\cdots+x_n)^{\ell(w)}\equiv
\prod_\alpha(x_1^{2^\alpha}+x_2^{2^\alpha}+\cdots+x_n^{2^\alpha})\mod
2.
\end{equation}
Hence multiplicity free monomials with {\it odd\,} coefficient can
appear only for permutation of length $\ell(w)\le 1$, i.e. for
$w=1$ or for transposition $w=(i,i+1)$. Note that
$$S_{(i,i+1)}(Z)=z_1+z_2+\cdots +z_i.$$
Therefore for even transposition $w=(i,i+1)$ coefficient
$S_{w}(1,1,\ldots, 1)=i$ in (\ref{mod2}) is even. So we proved
that $\varphi_T^*(\sigma_w)$ contains a monomial with odd
coefficient iff $w=(2j-1,2j)$ or $w=1$.
\end{proof}
Recall that identical substitution gives rise to basic marginal
inequality
\begin{equation}\label{basic_qubit}
\sum_i a_i(\rho_1^{(i)}-\rho_2^{(i)})\le \sum_{k}(\pm
a_1\pm a_2\pm\cdots\pm a_n)_k \rho_k
\end{equation}
for every extremal edge $a$. From Claim \ref{claim} it follows
\begin{thm}[modulo Conjecture \ref{conj}]\label{Mod_Qbt} All marginal inequaliteis for $n$
qubits can be obtained from basic ones $(\ref{basic_qubit})$ by
odd transposition $\rho_{2j-1}\leftrightarrow\rho_{2j}$ in the
right hand side,
accompanied with sign change
$a_i(\rho_1^{(i)}-\rho_2^{(i)})\mapsto-a_i(\rho_1^{(i)}-\rho_2^{(i)})
$ of one  term in the left hand side.
\end{thm}
\begin{rem}\label{Rem}
Right hand side of basic inequality (\ref{basic_qubit}) is
invariant under permutation of $a_i$. To get a strongest
inequality
we have to arrange them in the same order as quantities
$\delta_{i}=\rho^{(i)}_1-\rho^{(i)}_2$ to make LHS maximal
possible. Let's assume for certainty
\begin{equation}
\delta_{1}\le\delta_{2}\le\cdots \le
\delta_{n},\qquad
a_1\le a_2\le \cdots\le a_n.
\end{equation}
To get the strongest modified inequality one have to revert the
sign  of the minimal term $a_1\delta_1\mapsto -a_1\delta_1$ in the
LHS.
\end{rem}
\begin{Ex} {\it Three qubits.\,}
The above procedure being applied to extremal edges
(\ref{Edges3qbt}) for three qubits returns the following list of
marginal inequalities grouped by their extremal edges. The first
inequality in each group is the basic one. The transposed
eigenvalues in modified inequalities typeset in bold face. We
expect $\delta_i=\rho^{(i)}_1-\rho^{(i)}_2$ to be arranded in
increasing order $\delta_1\le\delta_2\le\delta_3$.
\begin{align*}
\delta_3
&\le\rho_1+\rho_2+\rho_3+\rho_4-\rho_5-\rho_6-\rho_7-\rho_8.\\[2mm]
\delta_2+\delta_3 &\le 2\rho_1+ 2\rho_2- 2\rho_7- 2\rho_8.\\[2mm]
\delta_1+\delta_2+\delta_3&\le3\rho_1+\rho_2+\rho_3+\rho_4-\rho_5-\rho_6-\rho_7-3\rho_8,\\
-\delta_1+\delta_2+\delta_3&\le3\boldsymbol{\rho_2}+\boldsymbol{\rho_1}+\rho_3+\rho_4-\rho_5-\rho_6-\rho_7-3\rho_8,\\
-\delta_1+\delta_2+\delta_3&\le3\rho_1+\rho_2+\rho_3+\rho_4-\rho_5-\rho_6-
\boldsymbol{\rho_8}-3\boldsymbol{\rho_7}.
\\[2mm]
\delta_1+\delta_2+2\delta_3&\le4\rho_1+2\rho_2+2\rho_3-2\rho_6-2\rho_7-4\rho_8,\\
-\delta_1+\delta_2+2\delta_3&\le4\boldsymbol{\rho_2}+2\boldsymbol{\rho_1}+2\rho_3-2\rho_6-2\rho_7-4\rho_8,\\
-\delta_1+\delta_2+2\delta_3&\le4\rho_1+2\rho_2+2\boldsymbol{\rho_4}-2\rho_6-2\rho_7-4\rho_8,\\
-\delta_1+\delta_2+2\delta_3&\le4\rho_1+2\rho_2+2\rho_3-2\boldsymbol{\rho_5}-2\rho_7-4\rho_8,\\
-\delta_1+\delta_2+2\delta_3&\le4\rho_1+2\rho_2+2\rho_3-2\rho_6-2\boldsymbol{\rho_8}-
4\boldsymbol{\rho_7}.
\end{align*}

One can check\footnote{Thanks to Maple simplex package.} that the
above  inequalities are independent, and other  inequalities
(\ref{QubitIneq}) follows from these ones, in conformity with
Conjecture \ref{conj}.
The later is still valid for four qubits,  however in this case
two modified inequalities are redundant, see Appendix. Actually
there are many ``trivial" redundant inequalities coming from
transposition of eigenvalues $\rho_{2j-1},\rho_{2j}$ entering in
RHS of basic inequality (\ref{basic_qubit}) with the same
coefficient. In settings of theorem \ref{ExtrEdge}, which deals
with extremal edges rather then cubicles, such transpositions are
forbidden and the ``trivial" redundancy never occurs.
\ref{Rem}.
\end{Ex}

\section{Representation theoretical
interpretation}\label{ReprTheor}

Apparently marginal inequalities give a complete solution of the
quantum marginal problem. This however may be an illusion, since
the number of inequalities increases drastically with rank of the
system. In this section we pursue another approach, based on
reduction
of the univariant marginal problem to {\it representation
theory\,} of the symmetric group. We recall first the basic facts
of the latter \cite{James,Macdonald,Sagan}.

\subsection{Digest of representation theory}\label{repres}

Let's start with ensemble of $N$ identical systems with state
space
\begin{equation}\label{tensor}\mathcal{H}^{\otimes
N}=\underbrace{\mathcal{H}\otimes\mathcal{H}\otimes\cdots\otimes\mathcal{H}}_N.
\end{equation}
Unitary $\SU(\mathcal{H})$ and symmetric ${\mathrm{S}}_N$ groups
act on this space from the left and from the right by formulae
\begin{align*}
g:\psi_1\otimes\psi_2\otimes\cdots\otimes\psi_N&\mapsto
g\psi_1\otimes g\psi_2\otimes\cdots\otimes g\psi_N,& g&\in \SU(\mathcal{H}),\\
\sigma:\psi_1\otimes\psi_2\otimes\cdots\otimes\psi_N&\mapsto
\psi_{\sigma(1)}\otimes\psi_{\sigma(2)}\otimes\cdots\otimes\psi_{\sigma(N)},
 &\sigma&\in {\mathrm{S}}_N.
\end{align*}
Issai~Schur in his celebrated thesis \cite{Schur} of 1901 found
decomposition of tensor space (\ref{tensor}) into irreducible
components with respect to these actions
\begin{equation}\label{Schur}
\mathcal{H}^{\otimes N}=\bigoplus_\lambda
\mathcal{H}_\lambda\otimes \mathcal{S}_\lambda,
\end{equation}
where $\mathcal{H}_\lambda$ and ${\mathcal{S}}_\lambda$ are
irreducible representations of the unitary  and the symmetric
 groups respectively. The components
 describe tensors of different types of symmetry. They are parameterized
 by {\it Young diagrams\,}
 $$\lambda:\lambda_1\ge\lambda_2\ge\cdots\ge\lambda_d\ge0,\qquad
 \lambda_1+\lambda_2+\cdots+\lambda_d=N$$ of $N$ cells and no more then
 $d=\dim\mathcal{H}$ rows of length $\lambda_i$. Representations
 $\mathcal{H}_\lambda$ and ${\mathcal{S}}_\lambda$ can be
 intrinsically characterized as follows.
\begin{itemize}
\item Let $V$ be irreducible representation of $\SU(\mathcal{H})$
and $\lambda$ be maximal diagram in {\it lexicographic order\,}
such that diagonal subgroup has eigenvector $\psi$ of weight
$\lambda$, that is
$$\mathrm{diag}(x_1,x_2,\ldots,x_d)\psi=
x_1^{\lambda_1}x_2^{\lambda_2}\cdots x_d^{\lambda_d}\psi.$$ Then
$V\simeq\mathcal{H}_\lambda$.
\item Let now $M$ be irreducible representation of $\mathrm{S}_N$,
and $\lambda$ be maximal diagram in the {\it majorization order\,}
(see $n^\circ$\ref{restr}) such that  $M$ contains a nonzero
invariant with respect to permutations in {\it rows\,} of diagram
$\lambda$ (filled in arbitrary way by numbers $1,2,\ldots,N$).
Then $M\simeq{\mathcal{S}}_\lambda$.
\end{itemize}
\begin{Ex} One row diagram $\lambda$ corresponds to trivial
representation
of the symmetric group. Representation $\mathcal{H}_\lambda$ in
this case consists of {\it symmetric tensors\,} in
$\mathcal{H}^{\otimes N}$ (Bose--Einstein statistics).

Column diagram $\lambda$ defines sign representation
$\sigma\mapsto \mathrm{sgn}(\sigma)$ of $S_N$, while
$\mathcal{H}_\lambda$ is the space of {\it antisymmetric
tensors\,} in $\mathcal{H}^{\otimes N}$ (Fermi--Dirak statistics).

Other diagrams correspond to more complicate symmetry types of
tensors which some physicists associate with {\it
parastatistics\,}.
\end{Ex}
\subsection{Degression: Hermitian spectral problem}\label{AddSpec}
Here we consider a {\it model example\,} which illustrates
relation between representation theory and spectral problems. As
we have seen at the end of $n^\circ$\ref{margin} univariant
marginal problem falls into this category.

Let's start with decomposition of tensor product of irreducible
representations of the unitary group $\SU(\mathcal{H})$
\begin{equation}\label{L-R}\mathcal{H}_\lambda\otimes\mathcal{H}_\mu=\sum_\nu
C_{\lambda\mu}^\nu\mathcal{H}_\nu.\end{equation} Representation
$\mathcal{H}_\nu$ enters into the decomposition with multiplicity
 $C_{\lambda\mu}^\nu$
which physicists  call {\it Clebsch--Gordon\,} and mathematiciens
{\it Littlewood--Richardson\,} coefficient. The last authors
invented in 1934 an efficient algorithm  for calculation
$C_{\lambda,\mu}^\nu$, known as {\it Littlewood--Richardson rule}
\footnote{ {\it ``The Littlewood--Richardson rule helped to get men on the moon,
but it was not proved until after they had got there."}, see
 \cite{James87}.}, see \cite{L-R,Macdonald} for details.
Littlewood--Richardson calculator is available at \cite{Buch}.

\begin{thm*}[Klyachko \cite{Kl98}]
The following conditions on Young dyagrams $\lambda,\mu,\nu$ are
equivalent
\begin{enumerate}
\item $C_{\lambda\mu}^\nu\ne0$, i.e.
$\mathcal{H}_\nu\subset\mathcal{H}_\lambda\otimes\mathcal{H}_\mu$.
\item There exist Hermitian operators $A,B,C=A+B$ in $\mathcal{H}$ with spectra
$\lambda,\mu,\nu$.
\end{enumerate}
Here we identify Young diagram $\lambda$ with integeral spectrum
$\lambda_1\ge\lambda_2\ge\lambda_3\ge\cdots$ formed by lengths of
its rows.
\end{thm*}
\begin{rem}\label{H-K}
(i) The conditions of the theorem are actually equivalent to a
system of linear inequalities on $\lambda,\mu,\nu$, known as {\it
Horn--Klyachko inequalities\,} \cite{Kl98,Ful2}.

(ii) The theorem as it stated  can be applied to integral spectra
only. However by scaling one can extend it to rational spectra,
and by limit arguments to real spectra as well. In the last case
we lose connection with representation theory, and compatibility
conditions for spectra of matrices $A,B,C=A+B$ are given by H-K
inequalities.
\end{rem}
\subsection{Back to quantum marginal problem} \label{main} Let now
consider tensor product of irreducible representations of the
symmetric group $S_N$
\begin{equation}\label{Kron}
\mathcal{S}_\lambda\otimes\mathcal{S}_\mu=\sum_\nu
g(\lambda,\mu,\nu)\mathcal{S}_\nu.
\end{equation}
The multiplicities $g(\lambda,\mu,\nu)$ in this case called {\it
Kronecker coefficients\,}.
In contrast
with Littlewood--Richardson ones
$C_{\lambda\mu}^\nu$ they
are symmetric in $\lambda,\mu,\nu$ and
\begin{equation}g(\lambda,\mu,\nu)=\dim(\mathcal{S}_\lambda\otimes\mathcal{S}_\mu\otimes
\mathcal{S}_\nu)^{S_N}.\end{equation}
The superscript refers to space of invariants $V^{S_N}=\{\,x\in
V\mid  x^\sigma=x \quad\forall \sigma\in S_N\,\}$.

Now we can state our main result for univariant marginal problem,
which is similar in form to the previous theorem on spectra of
Hermitian matrices $A$, $B$, and $A+B$.

\begin{thm}\label{mainThm} The following conditions are equivalent
\begin{enumerate}
\item $g(m\lambda,m\nu,m\mu)\ne0$ for some $m>0$, i.e.
$\mathcal{S}_{m\nu}\subset\mathcal{S}_{m\lambda}\otimes\mathcal{S}_{m\mu}$.
\item There exists mixed state of $\rho_{AB}$ of two component
system $\mathcal{H}_A\otimes\mathcal{H}_B$ with spectrum $\nu$ and
margins of spectra $\lambda,\mu$.
\item There exists pure state
$\psi\in\mathcal{H}_A\otimes\mathcal{H}_B\otimes\mathcal{H}_C$
with univariant margins of spectra $\lambda,\mu,\nu$.
\end{enumerate}
\end{thm}

\begin{proof}
The proof is a combination of well known results and runs as
follows.

$\bullet$ As we've yet mentioned in the proof of Theorem
\ref{cubicleThm} the quantum marginal problem amounts to
decomposition of projection of a coadjoint orbit of group
$\SU(\mathcal{H}_A\otimes\mathcal{H}_B)$ into coadjoint orbits of
subgroup $\SU(\mathcal{H}_A)\times\SU(\mathcal{H}_B)$.

$\bullet$ Combining this with Heckman's theorem \cite{Heck}, see
also
in \cite[Thm 3.4.2]{Ber-Sja}, we arrive at characterization of
quantum margins
by inclusion
\begin{equation}
\mathcal{H}_A^\lambda\otimes\mathcal{H}_B^\mu
\subset
\mathcal{H}_{AB}^\nu\left.\right|_{\SU(\mathcal{H}_A)\times\SU(\mathcal{H}_B)}
\end{equation}
for some spectra $\lambda,\mu,\nu$ proportional to
$\Spec\rho_A,\Spec\rho_B,\Spec\rho_{AB}$. The later, without loss
of generality, expected to be rational.

$\bullet$ Finally, the following equation for multiplicities
\begin{equation}\label{mult}\text{Mult. }\mathcal{H}_A^\mu\otimes\mathcal{H}_B^\nu
\text{ in }
\mathcal{H}_{AB}^\lambda\left.\right|_{\SU(\mathcal{H}_A)\times\SU(\mathcal{H}_B)}=
\text{Mult. }
 \mathcal{S}^\lambda\text{ in }
\mathcal{S}^\mu\otimes\mathcal{S}^\nu
\end{equation}
reduces the problem to Kronecker coefficients of the symmetric
group.

$\bullet$ Equivalence $(2)\Leftrightarrow(3)$ is strightforward
and independent of representation theory.
\end{proof}

In the rest of this paper we'll use the theorem in both directions
to obtain new results in quantum marginal problem and in
representation theory.

\section{Examples and applications}
%
\subsection{Polygonal inequalities revisited}\label{murn}


Admitting an abuse of language we'll use the same notation for
Young diagram $\lambda$ and  for the corresponding irreducible
representation of the symmetric group $S_n$. It is convenient to
single out the first row of diagram
$\lambda=(n-|\bar\lambda|,\bar\lambda)$. The remaining part
$\bar\lambda$ is called {\it reduced diagram}, and the number of
its cells $d(\lambda)=|\bar\lambda|$ is said to be {\it depth\,}
of $\lambda$.
We treat $n$ as a parameter and use parentheses to return the
original diagram $
(\bar\lambda)=(n-|\bar\lambda|,\bar\lambda)=\lambda$.

For small values of $n$ entries of $(\mu)=(n-|\mu|,\mu)$ may be
not in decreasing order. Then
the following rule is applied
\begin{equation}\label{rule}(\ldots,p,q,\ldots)=-(\ldots,q-1,p+1,\ldots)\end{equation}
 to transform
$(\mu)$ into a  Young diagram with sign $\pm$ or into zero (for
$q=p+1$). By the above convention $(\mu)$ is also understood  as a
{\it virtual representation\,} of $S_n$.

\begin{thm*}[Murnaghan \cite{Murnaghan38,Murnaghan55}, Littlewood \cite{Littlewood}]
$(1)$ Coefficients of decomposition
\begin{equation}\label{reduced}(\bar\lambda)\otimes(\bar\mu)=\sum_{\bar\nu}\bar
g(\bar\lambda,\bar\mu,\bar\nu)(\bar\nu).\end{equation} are
independent of $n$, and $g(\lambda,\mu,\nu)=\bar
g(\bar\lambda,\bar\mu,\bar\nu)$ for $n\gg 1$.

$(2)$ The coefficient $g(\lambda,\mu,\nu)$  vanishes except depth
of the diagrams
satisfies triangle inequalities
\begin{equation}\label{tri}d(\lambda)\le d(\mu)+d(\nu),\qquad
d(\mu)\le d(\nu)+d(\lambda),\qquad d(\nu)\le d(\lambda)+d(\mu).
\end{equation}

$(3)$ In the case of
equality in (\ref{tri}) the Kronecker coefficient coincides with
Littlewood-Richardson one for reduced dyagrams
\begin{equation}\label{bound} g(\lambda,\mu,\nu)=C_{\bar\mu\bar\nu}^{\bar\lambda},
\qquad d(\lambda)=d(\mu)+d(\nu).
\end{equation}
\end{thm*}
\begin{rem} Murnaghan \cite{Murnaghan38} gave dozens
examples of decomposition
(\ref{reduced}) like the following one
\begin{equation*}
\begin{split}
(n-2,2)\otimes(n-2,1^2)& =(n-1,1)+(n-2,2)+2(n-2,1^2)+(n-3,3)
\\&  +2(n-3,2,1)+ (n-3,1^3)+(n-4,3,1)+(n-4,2,1^2).
\end{split}
\end{equation*}
This equation literally gives Kronecker coefficients
for $n\ge7$. Otherwise rule (\ref{rule}) should be invoked which
may result in cancelation of some virtual components. See
\cite{Vallejo99} for a general stabilization bound for $n$.
\end{rem}
\begin{rem} Murnaghan theorem allows to defined new product of
Young diagrams
\begin{equation}
\bar{\lambda}\circ\bar{\mu}=\sum_{\bar{\nu}}q^{\bar{\lambda}+\bar{\mu}-\bar{\nu}}
g(\bar{\lambda},\bar{\mu},\bar{\nu})\bar{\nu}
\end{equation}
depending on parameter $q$. By (\ref{bound}) the multiplication is
a {\it deformation\,} of  cohomology ring of (infinite)
Grassmannian. The deformation is different from {\it quantum
cohomology\,} of Grassmannian which plays central role in unitary
spectral problem \cite{A-W}. It would be very interesting to find
a geometric interpretation of this deformation.
\end{rem}

Theorem \ref{mainThm} allows recast the second claim of
Murnaghan's theorem into the following result.

\begin{cor} Let
$\psi\in\mathcal{H}_A\otimes\mathcal{H}_B\otimes\mathcal{H}_C$ be
pure state with univariant margins of spectra $\lambda,\mu, \nu$.
Then their depth satisfies triangle inequalities
\begin{equation}\label{triang} d(\lambda)\le d(\mu)+d(\nu),\qquad
d(\mu)\le d(\nu)+d(\lambda),\qquad d(\nu)\le d(\lambda)+d(\mu),
\end{equation}
where the  depth of spectrum
$\lambda_1\ge\lambda_2\ge\lambda_3\ge\cdots$ is\,
 $\lambda_2+\lambda_3+\cdots:=d(\lambda)$.
\end{cor}

\begin{rem} Extension of the corollary to multicomponent systems
is straightforward. For $N$-qubit it returns polygonal
inequalities (\ref{poly}). The proof given by S.~Bravyi
\cite{Brav} actually works for depth as well.
\end{rem}

\subsection{Quasiclassical limit} Here we analyze the boundary
case (3) of Murnaghan's theorem in terms of the marginal problem.

Let $\rho_{AB}$ be mixed state of composite system
$\mathcal{H}_{AB}=\mathcal{H}_A\otimes\mathcal{H}_B$ with margins
$\rho_A,\rho_B$ and spectra
$\lambda_{AB},\lambda_{A},\lambda_{B}$. Suppose triangle
inequalities  for depth (\ref{triang}) degenerate into equality
\begin{equation}\label{deg}
d(\lambda_{AB})=d(\lambda_{A})+d(\lambda_{B}).
\end{equation}
Using notations
$$\psi_{AB},\qquad\psi_{A},\qquad\psi_{B}$$
for eigenstates of $\rho_{AB},\rho_{A},\rho_{B}$ with maximal
eigenvalues,
$$\overline{\mathcal{H}}_{AB},\qquad\overline{\mathcal{H}}_{A},\qquad
\overline{\mathcal{H}}_{B}$$
for their orthogonal complements, and
$$\overline{\rho}_{AB},\qquad\overline{\rho}_{A},\qquad\overline{\rho}_{B}$$
for restrictions of $\rho_{AB},\rho_A,\rho_{B}$ onto subspaces
$\overline{\mathcal{H}}_{AB},\overline{\mathcal{H}}_{A},\overline{\mathcal{H}}_{B}$,
one can show that equation (\ref{deg}) is equivalent to the
following conditions
\begin{itemize}
\item $\psi_{AB}=\psi_{A}\otimes \psi_{B}$;
\item $\overline{\rho}_{AB}$ has support in subspace
$\overline{\mathcal{H}}_A\otimes\psi_B+\psi_A\otimes\overline{\mathcal{H}}_B
\simeq\overline{\mathcal{H}}_A\oplus\overline{\mathcal{H}}_B$;
\item $\overline{\rho}_A$ and $\overline{\rho}_B$ are just
restrictions of $\overline{\rho}_{AB}$ onto
$\overline{\mathcal{H}}_A\otimes\psi_B$ and
$\psi_A\otimes\overline{\mathcal{H}}_B$,
\begin{equation}
\overline{\rho}_{AB}=\left[
\begin{tabular}{c|c}
$\overline{\rho}_A$& $\ast$\\
\hline
$\ast$&$\overline{\rho}_B$
\end{tabular}\right]
\end{equation}
\end{itemize}
We call this {\it quasiclassical limit\,} because all the events
happen in classical direct sum
$\overline{\mathcal{H}}_A\oplus\overline{\mathcal{H}}_B$ rather
then in quantum tensor product. In this case Theorem \ref{mainThm}
combined with equation (\ref{bound}) and Remark \ref{H-K} gives
\begin{thm}\label{quasi} The following conditions are equivalent
\begin{enumerate}
\item $C_{\lambda\mu}^\nu\ne 0$, that is
$\mathcal{H}_\nu\subset\mathcal{H}_\lambda\otimes\mathcal{H}_\mu$.
\item There exist Hermitian matrices $H_A$, $H_B$, and
$$
H_{AB}=\left[
\begin{tabular}{c|c}
$H_A$& $\ast$\\
\hline
$\ast$&$H_B$
\end{tabular}\right]
$$
 with spectra $\lambda$, $\mu$, and $\nu$ respectively.
 \item Horn--Klyachko inequalities hold for $\lambda,\mu,\nu$.
\end{enumerate}
\end{thm}
One have to use multiple L-R coefficients
$$\mathcal{H}_\lambda\otimes\mathcal{H}_\mu\otimes\mathcal{H}_\nu\otimes\cdots=
\sum_\sigma
C_{\lambda\mu\nu\cdots}^\sigma\mathcal{H}_\sigma$$ to deal with
arbitrary  number of diagonal blocks
$\mathcal{H}_A,\mathcal{H}_B,\mathcal{H}_C,\ldots$ of spectra
$\lambda,\mu,\nu,\ldots$. For blocks of size one the theorem
amounts to {\it Horn  majorization inequality\,} $$\Spec H \succ
\diag H$$
since for one row diagrams $a,b,c,\ldots$\quad
$C_{abc\cdots}^\sigma\neq0\Leftrightarrow \sigma\succ
(a,b,c,\ldots)$.

\begin{rem} It is more natural and easy deduce Theorem
\ref{quasi} directly from Berensein-Sjamaar theorem
\cite{Ber-Sja}, then deduce it from Murnaghan results. See
\cite{Li-Poon03} for another approach.
Note also that one can control spectrum of Hermitian matrix
$$
H=\left[
\begin{tabular}{c|c}
$\ast$& $X$\\
\hline
$X^*$&$\ast$
\end{tabular}\right]
$$
by linear inequalities in $\Spec H$ and {\it singular\,} spectrum
of $X$ \cite{Li-Poon,Fom-Ful}. It would be very interesting  to
merge these two results
to gain control over spectrum of Hamiltonian
$$
H_{AB}=\left[
\begin{tabular}{c|c}
$H_A$& $H_\text{int}$\\
\hline
$H_\text{int}^*$&$H_B$
\end{tabular}\right]
$$
of composite classical system in terms of spectra of the
components $A$, $B$, and singular spectrum of $H_\text{int}$,
which measures  the {\it strength\,} of their interaction. This,
however, can't be done by {\it linear\,} inequalities.

\end{rem}
\begin{rem} \label{sat} Note an important {\it saturation property\,} of
Littlewood--Richardson coefficients \cite{Knut-Tao}
$$ C_{m\lambda,m\mu}^{m\nu}\ne 0\Longleftrightarrow C_{\lambda\mu}^\nu\ne
0,$$ which simplifies the statement of Theorem \ref{quasi} (cf.
Theorem \ref{mainThm}).
It has no analogue  for Kronecker coefficients,
e.g. $2^2\subset2^2\otimes2^2$ but $1^2\not\subset1^2\otimes1^2$.
\begin{conj}
Saturation property still holds for reduced Kronecker coefficients
(\ref{reduced}).
\end{conj}
For example, the saturation conjecture is valid for
{\it leading\,} reduced coefficient
$\bar{g}(\bar{\lambda},\bar{\mu},\bar{\nu})=
C_{\bar{\mu}\bar{\nu}}^{\bar{\lambda}}$,
$\bar{\mu}+\bar{\nu}=\bar{\lambda}$, see Murnaghan theorem
$n^\circ$\ref{murn}.
\end{rem}

\subsection{Maximal eigenvalue of a state with given margins}
Here and in the next $n^\circ$\ref{rank} we consider the following
\begin{prob} How close to a pure state can be mixed state $\rho_{AB}$
with given margins $\rho_A,\rho_B$?
\end{prob}
As we have seen
in $n^\circ$\ref{pure} margins of a pure state are isospectral.
Hence for $\Spec\rho_A\neq\Spec\rho_B$ state $\rho_{AB}$ can't be
pure, and we want to get it as close to a pure state as possible.
Recall that state $\rho$ is pure iff its maximal eigenvalue is
equal to one. Hence the maximal eigenvalue, known as {\it spectral
norm\,} $\norm{\rho}_s$ of operator $\rho$,
may be considered as a measure of purity.
\begin{thm} \label{Max} Let $\rho_A,\rho_B$ be marginal states of spectra
$$\lambda_1\ge\lambda_2\ge\cdots\ge\lambda_n,\qquad\mu_1\ge\mu_2\ge\cdots\ge\mu_n$$
extend by zeros to make them of the same length. Then maximal
spectral norm of states $\rho_{AB}$ with margins $\rho_A,\rho_B$
is given by equation
\begin{equation}\label{max}
\max\norm{\rho_{AB}}_s=\sum_i\min(\lambda_i,\mu_i)=1-\frac12\sum_i\abs{\lambda_i-\mu_i}.
\end{equation}
\end{thm}

The proof amounts to application of Theorem \ref{mainThm} to the
following representation theoretical counterpart.
\begin{thm*}[Klemm \cite{Klemm}, Dvir \cite{Dvir}, Clausen \& Meier \cite{Cl-Ma}] Maximal
length of the first row of Young diagrams
$\nu\subset\lambda\otimes\mu$
is equal to $\abs{\lambda\cap\mu}=\sum_i\min(\lambda_i,\mu_i)$.
\end{thm*}

It is  an interesting problem to describe states $\rho_{AB}$ with
maximal spectral norm.
There is a simple formula for Kronecker coefficient of
representation $\nu\subset\lambda\otimes\mu$ with maximal first
row
\cite{Dvir,Cl-Ma}
\begin{equation}\label{DCM}
g(\lambda,\mu,\nu)=\sum_{\alpha,\beta}C_{\lambda\cap\mu,\alpha}^\lambda
C_{\lambda\cap\mu,\beta}^\mu\,
g(\alpha,\beta,\overline{\nu}),\qquad \nu_1=\abs{\lambda\cap\mu},
\end{equation}
which allows, in principle, recursively find maximal component
$\nu\subset\lambda\otimes\mu$ in lexicographic order. For example,
$\nu_2$ is equal to maximum of $\abs{\alpha\cap\beta}$ s.t. both
Littlewood--Richardson coefficients
$C_{\lambda\cap\mu,\alpha}^\lambda, C_{\lambda\cap\mu,\beta}^\mu$
are nonzero. This combinatorial problem amounts to construction of
a couple of skew standard tableaux of shapes $\lambda\backslash
(\lambda\cap\mu)$ and $\mu\backslash (\lambda\cap\mu)$ with as
close content as possible. I skip transposition of this into the
marginal problem key.

See \cite{Vallejo03} for construction of other components in
$\lambda\otimes\mu$.

\subsection{Rank of a state with given
margins}\label{rank} Pure state can be also characterized by its
rank
$$\rho \text{ pure}\Longleftrightarrow\rk\rho=1.$$
Hence $\rk\rho$ or better $\delta(\rho)=\log\rk\rho$ is another
measure of purity (or rather impurity). The later is just {\it
informational capacity\,} of support of $\rho$. It follows from
Example \ref{slice} that margins of a pure state satisfy triangle
inequalities
\begin{equation*}\label{triangle}\delta(\rho_A)\le\delta(\rho_B)+\delta(\rho_C),\quad
\delta(\rho_B)\le\delta(\rho_C)+\delta(\rho_A),\quad
\delta(\rho_C)\le\delta(\rho_A)+\delta(\rho_B),
\end{equation*}
and the same is true for $\delta(\rho_{A})$, $\delta(\rho_{B})$,
and $\delta(\rho_{AB})$. Representation theoretical counterpart of
this
\begin{equation}\label{schur}
\height(\nu)\le\height(\lambda)\cdot\height(\mu)\quad\text{for}\quad
\nu\subset\lambda\otimes\mu
\end{equation}
comes back to Schur \cite{Schur}, see also
\cite{Regev80}. Here the {\it height\,} $\height(\nu)$ of diagram
$\nu$ is the number of its rows.
 Note that decomposable state
$\rho_{AB}=\rho_A\otimes\rho_B$ has margins $\rho_A,\rho_B$, and
{\it maximal\,} possible rank $\rk\rho_{AB}=\rk{\rho_A}\rk\rho_B$.

\begin{thm}\label{Rank}  There exists state $\rho_{AB}$ of two component system with
given margins $\rho_A,\rho_B$ and
\begin{equation}
\rk\rho_{AB}\le\max(\rk\rho_A,\rk\rho_B).
\end{equation}
In addition, such a state can be taken with  maximal possible
spectral norm, given by Theorem \ref{Max}.
\end{thm}

The proof of the theorem once again is just a translation into
marginal problem language of the following representation
theoretical counterpart.
\begin{thm*}[Berele \& Imbo \cite{Ber-Imb}] Let
$\lambda$ and $\mu$ be Young diagram of height $\le k$.  Then
tensor product $\lambda\otimes\mu$ contains a component $\nu$  of
height $\le k$.
\end{thm*}
Vallejo \cite{Vallejo03} gave an explicit construction of such a
component $\nu\subset\lambda\otimes\mu$
with maximal possible first row $\nu_1=\abs{\lambda\cap\mu}$.
\begin{conj}
State $\rho_{AB}$ with maximal lexicographical spectrum has
minimal rank among all states with given margins $\rho_A,\rho_B$.
\end{conj}

\subsection{Relation with classical marginal problem}\label{cl}
Triplet of spectra $\lambda,\mu,\nu$ is
said to be {\it quasiclassical\,} iff there exist $p_{ij}\ge0$
such that
\begin{equation}\label{classical}\lambda=\sum_i
p_{ij},\qquad \mu=\sum_j p_{ij},\qquad p\prec\nu,
\end{equation} where the
last condition means that content of matrix $p=[p_{ij}]$ being
arranged in decreasing order is majorized by $\nu$.
 If the content of matrix $p$ coincides with
$\nu$ the triplet is said to be {\it classical\,}, cf.
$n^\circ$\ref{restr}.
We borrow this definition, with a minor modification,  from Sergei
Bravyi \cite{Brav} who proved
part $(1)\Leftrightarrow(2)$ of the next theorem.
\begin{thm} The following conditions on spectra $\lambda,\mu,\nu$
are equivalent
\begin{enumerate}
\item Triplet $\lambda,\mu,\nu$ is quasiclassical.
\item There exists mixed state $\rho_{AB}$ such that
$$\Spec\rho_{AB}\prec\nu,\quad \Spec\rho_A=\lambda,\quad
\Spec\rho_B=\mu.$$
\item There exists mixed state $\widetilde{\rho}_{AB}$ such that
$$\Spec\widetilde{\rho}_{AB}=\nu,\quad \Spec\widetilde{\rho}_A\succ\lambda,\quad
\Spec\widetilde{\rho}_B\succ\mu.$$
\end{enumerate}
\end{thm}
The second part $(1)\Leftrightarrow(3)$ comes from Theorem
\ref{mainThm} and  standard facts from representation theory of
the symmetric group:
\begin{itemize}
\item Let $[\lambda]$ be {\it permutation representation\,} induced from Schur
subgroup
$\mathrm{S}_\lambda=\mathrm{S}_{\lambda_1}\times\mathrm{S}_{\lambda_2}\times\cdots
\times\mathrm{S}_{\lambda_m}\subset\mathrm{S}_{n}$. Then
$\widetilde{\lambda}\subset
[\lambda]\Longleftrightarrow\widetilde{\lambda}\succ\lambda$.
\item Hence $\nu\subset\widetilde{\lambda}\otimes \widetilde{\mu}$
for some $\widetilde{\lambda}\succ\lambda$ and
$\widetilde{\nu}\succ\nu$ iff $\nu\subset[\lambda]\otimes[\mu]$.
\item
$\nu\subset[\lambda]\otimes[\mu]\Longleftrightarrow \lambda,\mu
\text{ are margins of integer matrix }p_{ij}\ge0\text{ s.t. }
p\prec\nu.$
\end{itemize}
The first property is actually a source of the definition of
irreducible representation $\mathcal{S}_{\lambda}$ (currently
shortened to $\lambda$) in $n^\circ$\ref{repres}.  The last one
can be seen from decomposition of tensor product
$[\lambda]\otimes[\mu]$ into direct sum of permutation modules
corresponding to intersection $Z_{ij}=X_i\cap Y_j$ of partitions
\begin{align*}\{\,1,2,\ldots,n\,\}&=\bigsqcup_i X_i,\qquad |X_i|=\lambda_i,\\
\{\,1,2,\ldots,n\,\}&=\bigsqcup_j Y_j,\qquad |Y_j|=\mu_j,
\end{align*}
of types $\lambda$ and $\mu$. It follows that  every component
$\nu\subset[\lambda]\otimes[\mu]$ satisfies (\ref{classical}) with
$p_{ij}=\abs{Z_{ij}}$, and vice versa.
\begin{cor}\label{minimal} Let $\rho_{AB}$ be mixed state with given margins
$\rho_A,\rho_B$ of spectra $\lambda_A,\lambda_B$. Suppose that
$\lambda_{AB}=\Spec\rho_{AB}$ is minimal possible in majorization
order. Then triplet $\lambda_{A},\lambda_{B},\lambda_{AB}$ is
classical.
\end{cor}

\begin{rem}
Implication $(3)\Rightarrow(2)$ has a simple analytical proof.
Indeed, inequality
$\Spec\widetilde{\rho}_A\succ\lambda$ means that
$\rho_A=\sum_ip_iU_i\widetilde{\rho}_AU_i^\dag$ has spectrum
$\lambda$ for some unitary operators $U_i$ and probabilities
$p_i\ge 0$ \cite{Wehrl}. In a similar way there exists
$\rho_B=\sum_jq_jV_j\widetilde{\rho}_BV_j^\dag$ with spectrum
$\mu$. Then
$$\rho_{AB}=\sum_{ij}p_iq_j(U_i\otimes V_j)\widetilde{\rho}_{AB}(U_i^\dag\otimes
V_j^\dag)$$ has margins $\rho_A,\rho_B$ and $\Spec
\rho_{AB}\prec\Spec\widetilde{\rho}_{AB}=\nu$.

Neither {\it analytical\,} proof of $(2)\Rightarrow(3)$,
nor {\it representation theoretical\,} proof of
$(3)\Rightarrow(2)$ are known, see $n^\circ$\ref{dual} below.
\end{rem}

\subsection{Two qubit revisited}\label{QbtRev} Here we return back to two qubit marginal problem
$n^\circ$\ref{two_qubit}
from representation theoretical
perspective. By Theorem \ref{mainThm} it amounts to
decomposition of tensor product
$$\lambda\otimes\mu=\sum_\nu g(\lambda,\mu,\nu)\nu
$$ of representations defined by
{\it two row\,} diagrams $\lambda=(\lambda_1,\lambda_2)$ and
$\mu=(\mu_1,\mu_2)$.  By Schur inequality (\ref{schur})
the components $\nu=(\nu_1,\nu_2,\nu_3,\nu_4)$ have at most four
rows.

Currently neither combinatorial description nor efficient
algorithm are available for Kronecker coefficients
$g(\lambda,\mu,\nu)$, in contrast with Littlewood--Richardson
ones. One of the first results in this direction belongs to
A.~Lascoux \cite{Las} who resolved  the case of hook diagrams
$\lambda,\mu$.
For two row diagrams the problem was first addressed by Remmel and
Whitehead in long paper \cite{Rem-Whd}, and later by Rosas
\cite{Rosas}. Unfortunately the results are too complicated to be
reproduced here. Mersedes Rosas
write down $g(\lambda,\mu,\nu)$ as a {\it difference\,} of the
number of lattice points in two polygons. A close look at her
formula shows that for symmetric group $\mathrm{S}_n$ of {\it odd
degree\,} $n$ one polygon can be moved into another by an affine
transformation respecting the lattice. As result we get
\begin{claim} For odd $n$ Kronecker coefficient
$g(\lambda,\mu,\nu)$ for two row diagrams $\lambda,\mu$ is equal
to the number of integers $a,b$ satisfying the following
conditions
\begin{subequations}\label{septagon}
\begin{align}
\nu_3+\nu_4&\le a \le \nu_2+\nu_4,\\
\nu_2+\nu_4&\le b \le\min(\nu_2+\nu_3,\nu_1+\nu_4),\\
\lambda_2-\mu_2&\le b-a \le\lambda_1-\mu_2,\label{mid}\\
a+b&\le \lambda_2+\mu_2,\\
a+b&\equiv \lambda_2+\mu_2 \pmod 2,
\end{align}
\end{subequations}
where we expect, without loss of generality, $\lambda_2\ge\mu_2$.
\end{claim}
It follows that
\begin{align*}
\lambda_2+\mu_2&\ge a+b \ge \nu_2+\nu_3+2\nu_4,\\
\abs{\lambda_2-\mu_2}&\le b-a \le \min(\nu_2-\nu_4,\nu_1-\nu_3),\\
\lambda_2 &\ge a\ge \nu_3+\nu_4,\\
\mu_2&\ge a\ge \nu_3+\nu_4,
\end{align*}
which in turn implies Bravyi inequalities (\ref{brav}). One can
easily check that the later are sufficient for
$g(\lambda,\mu,\nu)\ne0$. This gives another proof of Bravyi
theorem $n^\circ$\ref{two_qubit}.
\begin{rem}
For even $n$ a precise description of $g(\lambda,\mu,\nu)$ is more
complicate.
Admitting an absolute error $\le1$ the Kronecker coefficient  is
equal to {\it weighted\,} number of lattice points in heptagon
\eqref{septagon}, assigning weight $1/2$ to boundary points on
the line $b-a =\lambda_1-\mu_2$.


Available information on Kronecker coefficients of three row
diagrams seems to be insufficient to deduce Higichi theorem
$n^\circ$\ref{hig} or 397 independent marginal inequalities for
two qutrits in section \ref{3x3}.
\end{rem}

\section{Applications to representation theory} \label{Appl2Repr}
Unfortunately currently available information on Kronecker
coefficients is sparse and sporadic.\footnote{Arguably this is
{\it ``... the last major problem in ordinary representation
theory of $S_n$"} \cite{Dvir}.}
Therefore direct applications of Theorem \ref{mainThm} to marginal
problem is limited. We believe however that
interplay between 
representation theory of $\mathrm{S}_n$ and quantum margins
eventually may lead to  solution of both problems simultaneously,
as it happens, for example, for Hermitian spectral problem and
representations of $\SU(\mathcal{H})$. We leave this topic for a
future  study, and bound ourselves to few examples of using
Theorem \ref{mainThm} in backward direction.
One obstruction to this is lack of the saturation property (see
Remark \ref{sat}), which bounds rigorous results to
{\it stable regime} of very long diagrams $m\lambda$, $m\gg1$,
although some of them may still hold
in general.
We'll use special notation for {\it stable inclusion\,}
\begin{equation}\label{stable}\nu\underset{stbl}{\subset}
\lambda\otimes\mu \Longleftrightarrow m\nu\subset
m\lambda\otimes m\mu \text{ for some } m>0,
\end{equation}
where the diagrams are multiplied by $m$ row-wise.


\subsection{Stable support of Kronecker coefficients}
Here we are interested in triplets of Young diagrams
$\lambda,\mu,\nu$ such that  $\lambda\subset\mu\otimes\nu$, i.e.
in {\it support\,} of Kronecker coefficients. We have seen, for
example, that for such triplets $\height(\lambda)\le
\height(\mu)\height(\nu)$ ($n^\circ$\ref{rank}), and
$\abs{\overline{\lambda}}\le\abs{\overline{\mu}}+\abs{\overline{\nu}}$
($n^\circ$\ref{murn}).
It is much easier to describe {\it stable support\,},  defined by
stable inclusion \eqref{stable}.

\begin{thm}
Stable inclusion $\nu\underset{stab}{\subset} \lambda\otimes\mu$
for diagrams $\lambda,\mu$ of bounded heights
$$\height\lambda\le m,\qquad\height\mu\le n$$
is given by marginal inequalities for system of format $m\times n$
with $\lambda=\Spec \rho_A$, $\mu=\Spec \rho_B$, $\nu=\Spec
\rho_{AB}$, .
\end{thm}
\begin{proof}
This follows from Theorem \ref{mainThm} combined with marginal
inequalities of section \ref{MargIneq2}.\end{proof}

For example, for two row diagrams the stable inclusion amounts to
Bravyi inequalities, cf. section \ref{QbtRev}.

\begin{cor} Suppose $\nu^{(i)}\subset \lambda^{(i)}\otimes \mu^{(i)}$. Then
\begin{equation}\label{add}(\nu^{(1)}+{\nu}^{(2)})\underset{stab}{\subset}
(\lambda^{(1)}+{\lambda}^{(2)})\otimes(\mu^{(1)}+{\mu}^{(2)})
\end{equation}
where addition of the diagrams is defined row-wise.
\end{cor}
\begin{rem} Numerical experiments suggest that  inclusion \eqref{add}
actually holds in usual sense. This leads to the following
\end{rem}
\begin{conj} Triplets $\lambda,\mu,\nu$ s.t.
$g(\lambda,\mu,\nu)\ne 0$ form a semigroup with respect to
row-wise addition.
\end{conj}
One can expect that the semigroup is finitely generated if heights
of the diagrams are bounded.
\subsection{Rectangular
diagrams} In the case of {\it rectanguler\,} diagrams the previous
problem has very simple answer.
\begin{thm} Let $\lambda,\mu\nu$ be rectangular diagrams of heights
$l,m,n$. Then stable inclusion $\nu\underset{stab}{\subset}
\lambda\otimes\mu$ is equivalent to inequalities
$l\le mn$, $m\le nl$, $n\le lm$.
\end{thm}
\begin{proof} The theorem is about stable inclusion of trivial
representation in $\lambda\otimes\mu\otimes \nu$. By Theorem
\ref{mainThm} trivial representation corresponds to a {\it pure\,}
state $\rho_{ABC}$ of three component system $\mathcal{H}_A\otimes
\mathcal{H}_B\otimes
\mathcal{H}_C$ of formal $l\times m\times n$,
while rectangular diagrams $\lambda\mu\nu$ correspond to {\it
scalar\,} margins $\rho_A,\rho_B,\rho_C$. The result now follows
from criterion of existence of a pure state with scalar margins.
in section \ref{scalar}.
\end{proof}
\begin{rem} The simplest proof of the above criterion comes
from marginal inequalities of Theorem \ref{cubicleThm}.
\end{rem}

%
%
%
\appendix
\section{Marginal inequalities}

The appendix contains marginal inequalities for systems of rank
$\le 4$, i.e. for arrays  up to four qubits and for systems of
formats $2\times3$, $2\times4$, $3\times3$,  and
$2\times2\times3$. It covers all systems with few hundreds (rather
then thousands) marginal inequalities. Other statistical data are
collected in the following table.

\vspace{3mm}
\begin{center}
\begin{tabular}{|c|c|c|c|c|}
\hline
System & Rank&Inequalities&Edges &Permutations\\
\hline
$2\times2$&2&7\;(4)&3&9\\
\hline
$2\times2\times2$&3&40\;(10)&10&111\\
$2\times3$&3&41&6&98\\
\hline
$2\times4$&4&234&11&2191\\
$3\times3$&4&387\;(197)&17&2298\\
$2\times2\times3$&4&442\;(232)&39&3914\\
$2\times2\times2\times2$&4&805\;(50)&101&3723\\
\hline
\end{tabular}
\end{center}

For system of format $p\times q\times\cdots$ the last two columns
show the number of extremal edges and the number of permutations
$w\in S_{p\cdot q\cdot\ldots}$ of length $\ell(w)\le
\frac{p(p-1)}{2}+\frac{q(q-1)}{2}+\cdots$.
The number of marginal inequalities produced by permutation $w$
and extremal edge $E$ is equal to the number of components in
$\varphi_E^*(\sigma_w)$. For system of rank 4 this amounts all
together  to hundreds of thousands inequalities.
compared with seconds for rank $\le 3$). Note that the number of
inequalities in most cases can be essentially reduced using
symmetry w.r. to permutations of equidimensional components. These
numbers are shown in parenthesis. Further reduction is possible
using duality
$$(\lambda_1\ge\lambda_2\ge \cdots\ge\lambda_n)
\mapsto (-\lambda_n\ge-\lambda_{n-1}\ge \cdots\ge-\lambda_1)$$
applied to every spectrum involved in a marginal inequality.

Because of huge number of marginal inequalities for systems of
rank more then three, they provide only an illusion of a complete
solution of the
problem and
have limited practical value. I
reproduce the inequalities here primary as experimental data for
those brave people who may try to understand them better.

Details of the calculation \cite{Kl-Oz} will be published
elsewhere.
\subsection{Arrays of qubits}
In this case all marginal inequalities can be produced
automatically from the list of extremal edges, see $n^o$
\ref{mineq} for details.  All the spectra are arranged in
decreasing order $\rho_1\ge\rho_2\ge\cdots$. By technical reason
we normalize mixed states to trace zero. Hence univariant margins
have spectra $(\lambda,-\lambda),(\mu,-\mu),$ etc. To get marginal
inequalities in standard normalization $\Tr\rho=1$ one have to
change $2\lambda,2\mu,\ldots$ in LHS of the inequalities by
$\lambda_1-\lambda_2,\mu_1-\mu_2,\ldots$. To save space we skip
all the inequalities obtained from another one by a permutation of
qubits.
The reduced systems are complete if
marginal spectra arranged in increasing order
$\lambda\le\mu\le\cdots$.
\subsubsection{Two qubits}\label{2qbt} For completeness we reproduce here
Bravyi inequalities (\ref{brav}) for two qubits in current
notations.
\begin{align*}
2\lambda&\le\nu_1+\nu_2-\nu_3-\nu_4,\\
\lambda+\mu&\le \nu_1-\nu_4,\\
\lambda-\mu&\le\nu_2-\nu_4,\\
\lambda-\mu&\le\nu_1-\nu_3.
\end{align*}

\subsubsection{Three qubits}\label{3qbt} Below are 10 marginal inequalities
for three qubits.
By permutations of qubits they give a complete system of  40
independent constraints.
\begin{align*}
2\lambda &\le\rho_1+\rho_2+\rho_3+\rho_4-\rho_5-\rho_6-\rho_7-\rho_8,\\
\lambda+\mu &\le \rho_1+ \rho_2- \rho_7- \rho_8,\\[2mm]
2\lambda+\mu+\nu&\le2\rho_1+\rho_2+\rho_3-\rho_6-\rho_7-2\rho_8,\\
2\lambda+\mu-\nu&\le2\rho_2+\rho_1+\rho_3-\rho_6-\rho_7-2\rho_8,\\
2\lambda+\mu-\nu&\le2\rho_1+\rho_2+\rho_4-\rho_6-\rho_7-2\rho_8,\\
2\lambda+\mu-\nu&\le2\rho_1+\rho_2+\rho_3-\rho_5-\rho_7-2\rho_8,\\
2\lambda+\mu-\nu&\le2\rho_1+\rho_2+\rho_3-\rho_6-\rho_8-2\rho_7,\\[2mm]
2\lambda+2\mu+2\nu&\le3\rho_1+\rho_2+\rho_3+\rho_4-\rho_5-\rho_6-\rho_7-3\rho_8,\\
2\lambda+2\mu-2\nu&\le3\rho_2+\rho_1+\rho_3+\rho_4-\rho_5-\rho_6-\rho_7-3\rho_8,\\
2\lambda+2\mu-2\nu&\le3\rho_1+\rho_2+\rho_3+\rho_4-\rho_5-\rho_6-\rho_8-3\rho_7.
\end{align*}

\subsubsection{Four qubits}\label{4qbt}
In this case there are 805 independent marginal inequalities. To
save space we skip inequalities obtained from a preceding one by a
permutation of qubits. The resulting 50 inequalities are grouped
by the extremal edges. The basic inequality stands first, and the
remaining ones in the group are obtained by transposition of
eigenvalues $\tau_{2i-1},\tau_{2i}$ typeset in bold face and
change sign of the first coefficient in LHS, see Theorem
\ref{Mod_Qbt} for details. Note that in the fifth group odd
transpositions $\tau_{5},\tau_{6}$ and $\tau_{11},\tau_{12}$ give
redundant inequalities. Arguably this is the last system of
marginal constraints which can be published. For 5 qubits there
are more then thousand independent inequalities counted up to a
permutation of qubits.

{\footnotesize
\begin{align*}
2\rho &\le
\tau_1+\tau_2+\tau_3+\tau_4+\tau_5+\tau_6+\tau_7+\tau_8-\tau_9-\tau_{10}-
\tau_{11}-\tau_{12}-\tau_{13}-\tau_{14}-\tau_{15}-\tau_{16},\\[2mm]
2 \nu + 2 \rho &\le 2 \tau_1 + 2 \tau_2 + 2 \tau_3 + 2
\tau_4 - 2
\tau_{13} -
2 \tau_{14} - 2 \tau_{15} - 2 \tau_{16},\\[2mm]
2\mu+2\nu+2\rho &\le
3\tau_1+3\tau_2+\tau_3+\tau_4+\tau_5+\tau_6+\tau_7+\tau_8-\tau_9-\tau_{10}-\tau_{11}-
\tau_{12}-\tau_{13}-\tau_{14}-3\tau_{15}-3\tau_{16},\\[2mm]
2 \mu + 2 \nu + 4 \rho &\le 4 \tau_1 + 4 \tau_2 + 2 \tau_3 + 2
\tau_4 + 2
\tau_5 + 2 \tau_6 - 2 \tau_{11} - 2 \tau_{12} - 2 \tau_{13} - 2 \tau_{14} - 4 \tau_{15} -
4 \tau_{16},
\displaybreak[0]\\[2mm]
2 \lambda + 2 \mu + 2 \nu + 2 \rho&\le 4
\tau_1 + 2\tau_2 + 2\tau_3 + 2 \tau_4 + 2 \tau_5 - 2 \tau_{12} - 2
\tau_{13} - 2\tau_{14} - 2 \tau_{15} -
4\tau_{16},\\
2 \lambda + 2 \mu + 2 \nu - 2 \rho&\le 4 \boldsymbol{\tau_2} + 2
\boldsymbol{\tau_1} + 2
\tau_3 + 2 \tau_4 + 2 \tau_5 - 2 \tau_{12} - 2 \tau_{13} - 2
\tau_{14} - 2 \tau_{15} - 4
\tau_{16},\\
2 \lambda + 2 \mu + 2 \nu - 2 \rho &\le 4 \tau_1 + 2 \tau_2 + 2
\tau_3 +
2 \tau_4 + 2 \tau_5 - 2 \tau_{12} - 2 \tau_{13} - 2 \tau_{14} - 2
\boldsymbol{\tau_{16}} - 4\boldsymbol{\tau_{15}},
\displaybreak[0]\\[2mm]
2 \lambda + 2 \mu + 2 \nu + 4 \rho &\le 5 \tau_1 + 3 \tau_2 + 3
\tau_3 + 3 \tau_4 + \tau_5 + \tau_6 + \tau_7 +\tau_8-\tau_9 - \tau_{10} - \tau_{11} -
\tau_{12}
- 3 \tau_{13}- 3 \tau_{14} - 3 \tau_{15} - 5 \tau_{16},\\
2 \lambda + 2 \mu - 2 \nu + 4 \rho &\le 5 \boldsymbol{\tau_2} + 3
\boldsymbol{\tau_1} + 3
\tau_3 + 3 \tau_4 + \tau_5 + \tau_6 + \tau_7 + \tau_8 - \tau_9 -
\tau_{10} - \tau_{11} - \tau_{12}
- 3 \tau_{13} - 3 \tau_{14} - 3 \tau_{15} - 5 \tau_{16},\\
2 \lambda + 2 \mu - 2 \nu + 4 \rho &\le5 \tau_1 + 3 \tau_2 + 3
\tau_3 + 3 \tau_4 + \tau_5 + \tau_6 + \tau_7 + \tau_8 - \tau_9 -
\tau_{10} - \tau_{11} - \tau_{12}
- 3 \tau_{13} - 3 \tau_{14} - 3 \boldsymbol{\tau_{16}} - 5
\boldsymbol{\tau_{15}},
\displaybreak[0]\\[2mm]
2 \lambda + 2 \mu + 2 \nu + 6 \rho&\le 6 \tau_1 + 4 \tau_2 + 4
\tau_3 + 4
\tau_4 + 2 \tau_5 + 2 \tau_6 + 2 \tau_7 - 2 \tau_{10} - 2 \tau_{11} - 2 \tau_{12} - 4
\tau_{13} - 4 \tau_{14} - 4 \tau_{15} - 6 \tau_{16},\\
2 \lambda + 2 \mu - 2 \nu + 6 \rho &\le 6 \boldsymbol{\tau_2} + 4
\boldsymbol{\tau_1} + 4
\tau_3 + 4
\tau_4 + 2 \tau_5 + 2 \tau_6 + 2 \tau_7 - 2 \tau_{10} - 2 \tau_{11} - 2 \tau_{12} - 4
\tau_{13} - 4 \tau_{14} - 4 \tau_{15} - 6 \tau_{16},\\
2 \lambda + 2 \mu - 2 \nu + 6 \rho &\le 6 \tau_1 + 4 \tau_2 + 4
\tau_3 +
4 \tau_4 + 2 \tau_5 + 2 \tau_6 + 2 \boldsymbol{\tau_8} - 2
\tau_{10} - 2
\tau_{11}
 - 2 \tau_{12} -4 \tau_{13} - 4 \tau_{14} - 4 \tau_{15} - 6 \tau_{16},\\
 2 \lambda + 2 \mu - 2 \nu + 6 \rho&\le 6 \tau_1 + 4 \tau_2 + 4
\tau_3 + 4 \tau_4 + 2 \tau_5 + 2 \tau_6 + 2 \tau_7 - 2 \boldsymbol{\tau_9} - 2
\tau_{11} - 2 \tau_{12} - 4
\tau_{13} - 4 \tau_{14} - 4 \tau_{15} - 6 \tau_{16},\\
2 \lambda + 2 \mu - 2 \nu + 6 \rho &\le6 \tau_1 + 4 \tau_2 + 4
\tau_3 + 4 \tau_4 + 2 \tau_5 + 2 \tau_6 + 2 \tau_7 - 2 \tau_{10} -
2 \tau_{11} - 2 \tau_{12} - 4 \tau_{13} - 4 \tau_{14} - 4
\boldsymbol{\tau_{16}} - 6 \boldsymbol{\tau_{15}},
\displaybreak[0]\\[2mm]
2 \lambda + 2 \mu + 4 \nu + 4 \rho &\le 6 \tau_1 + 4 \tau_2 + 4
\tau_3 + 2 \tau_4 + 2 \tau_5 + 2 \tau_6 - 2 \tau_{11} - 2 \tau_{12} - 2 \tau_{13} - 4
\tau_{14} - 4 \tau_{15} - 6 \tau_{16},\\
2 \lambda - 2 \mu + 4 \nu + 4 \rho&\le 6 \boldsymbol{\tau_2} + 4
\boldsymbol{\tau_1} + 4
\tau_3 + 2 \tau_4 + 2 \tau_5 + 2 \tau_6 - 2 \tau_{11} - 2
\tau_{12} - 2 \tau_{13} - 4 \tau_{14} -
4 \tau_{15} - 6 \tau_{16},\\
2 \lambda - 2 \mu + 4 \nu + 4 \rho &\le 6 \tau_1 + 4 \tau_2 + 4
\boldsymbol{\tau_4} + 2 \boldsymbol{\tau_3} + 2 \tau_5 + 2 \tau_6 - 2 \tau_{11} - 2
\tau_{12} - 2 \tau_{13} - 4 \tau_{14} -
4 \tau_{15} - 6 \tau_{16},\\
2\lambda - 2 \mu + 4 \nu + 4 \rho &\le6 \tau_1 + 4 \tau_2 + 4
\tau_3 + 2
\tau_4 + 2 \tau_5 + 2 \tau_6 - 2 \tau_{11} - 2 \tau_{12} - 2 \boldsymbol{\tau_{14}} -
4 \boldsymbol{\tau_{13}} - 4
\tau_{15} - 6 \tau_{16},\\
2 \lambda - 2 \mu + 4 \nu + 4 \rho &\le 6 \tau_1 + 4 \tau_2 + 4
\tau_3 + 2
\tau_4 + 2 \tau_5 + 2 \tau_6 - 2 \tau_{11} - 2 \tau_{12} - 2 \tau_{13} - 4 \tau_{14} - 4
\boldsymbol{\tau_{16}} - 6 \boldsymbol{\tau_{15}},
\displaybreak[0]\\[2mm]
2 \lambda + 2 \mu + 4 \nu + 6 \rho &\le 7 \tau_1 + 5 \tau_2 + 5
\tau_3 + 3 \tau_4 + 3 \tau_5 + \tau_6 + \tau_7 + \tau_8 - \tau_9 -
\tau_{10} - \tau_{11} - 3
\tau_{12} - 3 \tau_{13} - 5 \tau_{14} - 5 \tau_{15} - 7 \tau_{16},\\
2 \lambda - 2 \mu + 4 \nu + 6 \rho &\le 7 \boldsymbol{\tau_2} + 5
\boldsymbol{\tau_1} + 5
\tau_3 + 3 \tau_4 + 3 \tau_5 + \tau_6 + \tau_7 + \tau_8 - \tau_9 -
\tau_{10} - \tau_{11} - 3
\tau_{12} - 3 \tau_{13} - 5 \tau_{14} - 5 \tau_{15} - 7 \tau_{16},\\
2 \lambda - 2 \mu + 4 \nu + 6 \rho &\le7 \tau_1 + 5 \tau_2 + 5
\boldsymbol{\tau_4} + 3 \boldsymbol{\tau_3} + 3 \tau_5 + \tau_6 + \tau_7 + \tau_8 - \tau_9 -
\tau_{10} - \tau_{11} - 3
\tau_{12} - 3 \tau_{13} - 5 \tau_{14} - 5 \tau_{15} - 7 \tau_{16},\\
2 \lambda - 2 \mu + 4 \nu + 6 \rho &\le7 \tau_1 + 5 \tau_2 + 5
\tau_3 + 3 \tau_4 + 3 \boldsymbol{\tau_6} + \boldsymbol{\tau_5} + \tau_7 + \tau_8 - \tau_9 -
\tau_{10} - \tau_{11} - 3
\tau_{12} - 3 \tau_{13} - 5 \tau_{14} - 5 \tau_{15} - 7 \tau_{16},\\
2 \lambda - 2 \mu + 4 \nu + 6 \rho &\le7 \tau_1 + 5 \tau_2 + 5
\tau_3 + 3 \tau_4 + 3 \tau_5 + \tau_6 + \tau_7 + \tau_8 - \tau_9 -
\tau_{10} - \boldsymbol{\tau_{12}} - 3
\boldsymbol{\tau_{11}} - 3 \tau_{13} - 5 \tau_{14} - 5 \tau_{15} - 7 \tau_{16},\\
2 \lambda - 2 \mu + 4 \nu + 6 \rho &\le7 \tau_1 + 5 \tau_2 + 5
\tau_3 + 3 \tau_4 + 3 \tau_5 + \tau_6 + \tau_7 + \tau_8 - \tau_9 -
\tau_{10} - \tau_{11} - 3
\tau_{12} - 3 \boldsymbol{\tau_{14}} - 5 \boldsymbol{\tau_{13}} - 5 \tau_{15} - 7 \tau_{16},\\
2 \lambda - 2 \mu + 4 \nu + 6 \rho &\le 7 \tau_1 + 5 \tau_2 + 5
\tau_3 + 3 \tau_4 + 3 \tau_5 + \tau_6 + \tau_7 + \tau_8 - \tau_9 -
\tau_{10} - \tau_{11} - 3
\tau_{12} - 3 \tau_{13} - 5 \tau_{14} - 5 \boldsymbol{\tau_{16}} - 7 \boldsymbol{\tau_{15}},
\displaybreak[0]\\[2mm]
2 \lambda + 4 \mu + 4 \nu + 6 \rho&\le 8 \tau_1 + 6 \tau_2 + 4
\tau_3 + 4 \tau_4 + 2 \tau_5 + 2 \tau_6 + 2 \tau_7 - 2 \tau_{10} -
2 \tau_{11} - 2 \tau_{12} -
4 \tau_{13} - 4 \tau_{14} - 6 \tau_{15} - 8 \tau_{16},\\
-2 \lambda + 4 \mu + 4 \nu + 6 \rho&\le 8 \boldsymbol{\tau_2} + 6
\boldsymbol{\tau_1} + 4
\tau_3 + 4 \tau_4 + 2 \tau_5 + 2 \tau_6 + 2 \tau_7 - 2 \tau_{10} -
2 \tau_{11} - 2 \tau_{12} -
4 \tau_{13} - 4 \tau_{14} - 6 \tau_{15} - 8 \tau_{16},\\
-2 \lambda + 4 \mu + 4 \nu + 6 \rho &\le8 \tau_1 + 6 \tau_2 + 4
\tau_3 + 4 \tau_4 + 2 \tau_5 + 2 \tau_6 + 2 \boldsymbol{\tau_8} - 2 \tau_{10} -
2 \tau_{11} - 2 \tau_{12} -
4 \tau_{13} - 4 \tau_{14} - 6 \tau_{15} - 8 \tau_{16},\\
-2 \lambda + 4 \mu + 4 \nu + 6 \rho&\le 8 \tau_1 + 6 \tau_2 + 4
\tau_3 + 4 \tau_4 + 2 \tau_5 + 2 \tau_6 + 2 \tau_7 - 2 \boldsymbol{\tau_9} - 2
\tau_{11} - 2 \tau_{12} - 4
\tau_{13} - 4 \tau_{14} - 6 \tau_{15} - 8 \tau_{16},\\
 -2 \lambda + 4 \mu + 4 \nu + 6 \rho &\le8
\tau_1 + 6 \tau_2 + 4
\tau_3 + 4 \tau_4 + 2 \tau_5 + 2 \tau_6 + 2 \tau_7 - 2 \tau_{10} -
2 \tau_{11} - 2 \tau_{12} - 4 \tau_{13} - 4 \tau_{14} - 6
\boldsymbol{\tau_{16}} - 8 \boldsymbol{\tau_{15}},
\displaybreak[0]\\[2mm]
2 \lambda + 2 \mu + 4 \nu + 8 \rho&\le 8 \tau_1 + 6 \tau_2 + 6
\tau_3 + 4 \tau_4 + 4 \tau_5 + 2 \tau_6 + 2 \tau_7 - 2 \tau_{10} -
2 \tau_{11} - 4 \tau_{12} -
4 \tau_{13} - 6 \tau_{14} - 6 \tau_{15} - 8 \tau_{16},\\
2 \lambda - 2 \mu + 4 \nu + 8 \rho&\le 8 \boldsymbol{\tau_2} + 6
\boldsymbol{\tau_1} + 6
\tau_3 + 4 \tau_4 + 4 \tau_5 + 2 \tau_6 + 2 \tau_7 - 2 \tau_{10} -
2 \tau_{11} - 4 \tau_{12} -
4 \tau_{13} - 6 \tau_{14} - 6 \tau_{15} - 8 \tau_{16},\\
2 \lambda - 2 \mu + 4 \nu + 8 \rho &\le 8 \tau_1 + 6 \tau_2 + 6
\boldsymbol{\tau_4} + 4
\boldsymbol{\tau_3} + 4 \tau_5 + 2 \tau_6 + 2 \tau_7 - 2 \tau_{10} - 2 \tau_{11} - 4 \tau_{12} - 4
\tau_{13} - 6 \tau_{14} - 6 \tau_{15} - 8 \tau_{16},\\
2 \lambda - 2 \mu + 4 \nu + 8 \rho &\le 8 \tau_1 + 6 \tau_2 + 6
\tau_3 + 4 \tau_4 + 4 \boldsymbol{\tau_6} + 2 \boldsymbol{\tau_5} + 2 \tau_7 - 2 \tau_{10} -
2 \tau_{11} - 4 \tau_{12} -
4 \tau_{13} - 6 \tau_{14} - 6 \tau_{15} - 8 \tau_{16},\\
2 \lambda - 2 \mu + 4 \nu + 8 \rho&\le 8 \tau_1 + 6 \tau_2 + 6
\tau_3 + 4 \tau_4 + 4 \tau_5 + 2 \tau_6 + 2 \boldsymbol{\tau_8} - 2 \tau_{10} -
2 \tau_{11} - 4 \tau_{12} -
4 \tau_{13} - 6 \tau_{14} - 6 \tau_{15} - 8 \tau_{16},\\
2 \lambda - 2 \mu + 4 \nu + 8 \rho&\le 8 \tau_1 + 6 \tau_2 + 6
\tau_3 + 4 \tau_4 + 4 \tau_5 + 2 \tau_6 + 2 \tau_7 - 2 \boldsymbol{\tau_9} - 2
\tau_{11} - 4 \tau_{12} - 4
\tau_{13} - 6 \tau_{14} - 6 \tau_{15} - 8 \tau_{16},\\
2 \lambda - 2 \mu + 4 \nu + 8 \rho&\le 8 \tau_1 + 6 \tau_2 + 6
\tau_3 + 4 \tau_4 + 4 \tau_5 + 2 \tau_6 + 2 \tau_7 - 2 \tau_{10} -
2 \boldsymbol{\tau_{12}} - 4 \boldsymbol{\tau_{11}} -
4 \tau_{13} - 6 \tau_{14} - 6 \tau_{15} - 8 \tau_{16},\\
2 \lambda - 2 \mu + 4 \nu + 8 \rho &\le8 \tau_1 + 6 \tau_2 + 6
\tau_3 + 4 \tau_4 + 4 \tau_5 + 2 \tau_6 + 2 \tau_7 - 2 \tau_{10} -
2 \tau_{11} - 4 \tau_{12} -
4 \boldsymbol{\tau_{14}} - 6 \boldsymbol{\tau_{13}} - 6 \tau_{15} - 8 \tau_{16},\\
2 \lambda - 2 \mu + 4 \nu + 8 \rho &\le8 \tau_1 + 6 \tau_2 + 6
\tau_3 + 4 \tau_4 + 4 \tau_5 + 2 \tau_6 + 2 \tau_7 - 2 \tau_{10} -
2 \tau_{11} - 4 \tau_{12} - 4 \tau_{13} - 6 \tau_{14} - 6
\boldsymbol{\tau_{16}} - 8
\boldsymbol{\tau_{15}},
\displaybreak[0]\\[2mm]
2 \lambda + 4 \mu + 6 \nu + 8 \rho&\le 10 \tau_1 + 8 \tau_2 + 6
\tau_3 + 4 \tau_4 + 4 \tau_5 + 2 \tau_6 + 2 \tau_7 - 2 \tau_{10} -
2 \tau_{11} - 4 \tau_{12} -
4 \tau_{13} - 6 \tau_{14} - 8 \tau_{15} - 10 \tau_{16},\\
-2 \lambda + 4 \mu + 6 \nu + 8 \rho &\le 10 \boldsymbol{\tau_2} +
8
\boldsymbol{\tau_1} + 6
\tau_3 + 4
\tau_4 + 4 \tau_5 + 2 \tau_6 + 2 \tau_7 - 2 \tau_{10} - 2 \tau_{11} - 4 \tau_{12} - 4
\tau_{13} - 6 \tau_{14} - 8 \tau_{15} - 10 \tau_{16},\\
-2 \lambda + 4 \mu + 6 \nu + 8 \rho &\le 10 \tau_1 + 8 \tau_2 + 6
\boldsymbol{\tau_4} + 4 \boldsymbol{\tau_3} + 4 \tau_5 + 2 \tau_6 + 2 \tau_7 - 2 \tau_{10} -
2 \tau_{11} - 4 \tau_{12}
- 4 \tau_{13} - 6 \tau_{14} - 8 \tau_{15} - 10 \tau_{16},\\
-2 \lambda + 4 \mu + 6 \nu + 8 \rho &\le 10 \tau_1 + 8 \tau_2 + 6
\tau_3 + 4
\tau_4 + 4 \boldsymbol{\tau_6} + 2 \boldsymbol{\tau_5} + 2 \tau_7 - 2 \tau_{10} - 2 \tau_{11} - 4 \tau_{12} - 4
\tau_{13} - 6 \tau_{14} - 8 \tau_{15} - 10 \tau_{16},\\
-2 \lambda + 4 \mu + 6 \nu + 8 \rho&\le 10 \tau_1 + 8 \tau_2 + 6
\tau_3 + 4 \tau_4 + 4 \tau_5 + 2 \tau_6 + 2 \boldsymbol{\tau_8} - 2 \tau_{10} -
2 \tau_{11} - 4 \tau_{12}
- 4 \tau_{13} - 6 \tau_{14} - 8 \tau_{15} - 10 \tau_{16},\\
-2 \lambda + 4 \mu + 6 \nu + 8 \rho &\le 10 \tau_1 + 8 \tau_2 + 6
\tau_3 + 4
\tau_4 + 4 \tau_5 + 2 \tau_6 + 2 \tau_7 - 2 \boldsymbol{\tau_9} - 2 \tau_{11} - 4 \tau_{12} - 4
\tau_{13} - 6 \tau_{14} - 8 \tau_{15} - 10 \tau_{16},\\
-2 \lambda + 4 \mu + 6 \nu + 8 \rho&\le 10 \tau_1 + 8 \tau_2 + 6
\tau_3 + 4 \tau_4 + 4 \tau_5 + 2 \tau_6 + 2 \tau_7 - 2 \tau_{10} -
2 \boldsymbol{\tau_{12}} - 4 \boldsymbol{\tau_{11}}
- 4 \tau_{13} - 6 \tau_{14} - 8 \tau_{15} - 10 \tau_{16},\\
-2 \lambda + 4 \mu + 6 \nu + 8 \rho &\le 10 \tau_1 + 8 \tau_2 + 6
\tau_3 + 4 \tau_4 + 4 \tau_5 + 2 \tau_6 + 2 \tau_7 - 2 \tau_{10} -
2 \tau_{11} - 4 \tau_{12}
- 4 \boldsymbol{\tau_{14}} - 6 \boldsymbol{\tau_{13}} - 8 \tau_{15} - 10 \tau_{16},\\
-2 \lambda + 4 \mu + 6 \nu + 8 \rho &\le 10 \tau_1 + 8 \tau_2 + 6
\tau_3 + 4 \tau_4 + 4 \tau_5 + 2 \tau_6 + 2 \tau_7 - 2 \tau_{10} -
2 \tau_{11} - 4 \tau_{12} - 4 \tau_{13} - 6 \tau_{14} - 8
\boldsymbol{\tau_{16}} - 10 \boldsymbol{\tau_{15}}.
\end{align*}}
\subsection{Two qutrits}\label{3x3}
Below are $197$ marginal inequalities for system
 of two qutrits $\mathcal{H}\otimes
\mathcal{H}$, $\dim\mathcal{H}=3$. Together with inequalities obtained
by transposition of qutrits they form a complete and independent
system of 387 marginal constraints. One can check that for mixed
state of rank three (i.e. for $\nu_i=0,i>3$) the system
amounts to Higuchi inequalities $n^o$ \ref{hig}. {\footnotesize

}

\subsection{Systems of format $2\times n$}
In this case all cubicles and extremal edges are explicitly known,
see Example \ref{2xn}.  The tables below give marginal
inequalities for $n=3,4$.
For $n>4$
the number of marginal inequalities increase  to thousands and
can't be published.
\subsubsection{Format $2\times3$}\label{2x3} In this case there are 41 independent
marginal inequalities. {\footnotesize
\begin{align*}
\mu_1-\mu_2 &\le \nu_1+\nu_2+\nu_3-\nu_4-\nu_5-\nu_6.
\displaybreak[0]\\[2mm]
\lambda_1+\lambda_2-2\lambda_3 &\le \nu_1+\nu_2+\nu_3+\nu_4-2\nu_5-2\nu_6,\\
\lambda_2+\lambda_3-2\lambda_1 &\le \nu_1+\nu_2+\nu_3+\nu_6-2\nu_4-2\nu_5.
\displaybreak[0]\\[2mm]
2\lambda_1-\lambda_2-\lambda_3 &\le 2\nu_1+2\nu_2-\nu_3-\nu_4-\nu_5-\nu_6,\\
2\lambda_3-\lambda_1-\lambda_2 &\le
2\nu_2+2\nu_3-\nu_1-\nu_4-\nu_5-\nu_6.
\displaybreak[0]\\[2mm]
2\lambda_1-2\lambda_3-\mu_2+\mu_1 &\le 3\nu_1+\nu_2+\nu_3-\nu_4-\nu_5-3\nu_6,\\
2\lambda_1-2\lambda_3+\mu_2-\mu_1 &\le 3\nu_2+\nu_1+\nu_3-\nu_4-\nu_5-3\nu_6,\\
2\lambda_2-2\lambda_3-\mu_2+\mu_1 &\le 3\nu_2+\nu_1+\nu_3-\nu_4-\nu_5-3\nu_6,\\
2\lambda_1-2\lambda_2+\mu_2-\mu_1 &\le 3\nu_2+\nu_1+\nu_4-\nu_3-\nu_5-3\nu_6,\\
2\lambda_1-2\lambda_3+\mu_2-\mu_1 &\le 3\nu_1+\nu_2+\nu_4-\nu_3-\nu_5-3\nu_6,\\
2\lambda_1-2\lambda_2-\mu_2+\mu_1 &\le 3\nu_1+\nu_2+\nu_4-\nu_3-\nu_5-3\nu_6,\\
2\lambda_2-2\lambda_3-\mu_2+\mu_1 &\le 3\nu_1+\nu_2+\nu_4-\nu_3-\nu_5-3\nu_6,\\
2\lambda_3-2\lambda_1-\mu_2+\mu_1 &\le 3\nu_3+\nu_1+\nu_4-\nu_2-\nu_5-3\nu_6,\\
2\lambda_3-2\lambda_1+\mu_2-\mu_1 &\le 3\nu_3+\nu_2+\nu_4-\nu_1-\nu_5-3\nu_6,\\
2\lambda_3-2\lambda_2+\mu_2-\mu_1 &\le 3\nu_2+\nu_3+\nu_4-\nu_1-\nu_5-3\nu_6,\\
2\lambda_3-2\lambda_1-\mu_2+\mu_1 &\le 3\nu_2+\nu_3+\nu_4-\nu_1-\nu_5-3\nu_6,\\
2\lambda_2-2\lambda_1+\mu_2-\mu_1 &\le 3\nu_1+\nu_2+\nu_6-\nu_4-\nu_3-3\nu_5,\\
2\lambda_3-2\lambda_1-\mu_2+\mu_1 &\le 3\nu_1+\nu_2+\nu_6-\nu_4-\nu_3-3\nu_5,\\
2\lambda_2-2\lambda_3+\mu_2-\mu_1 &\le 3\nu_1+\nu_2+\nu_4-\nu_3-\nu_6-3\nu_5,\\
2\lambda_1-2\lambda_2+\mu_2-\mu_1 &\le 3\nu_2+\nu_1+\nu_3-\nu_4-\nu_6-3\nu_5,\\
2\lambda_2-2\lambda_3+\mu_2-\mu_1 &\le 3\nu_2+\nu_1+\nu_3-\nu_4-\nu_6-3\nu_5,\\
2\lambda_1-2\lambda_3+\mu_2-\mu_1 &\le 3\nu_1+\nu_2+\nu_3-\nu_6-\nu_4-3\nu_5,\\
2\lambda_1-2\lambda_2-\mu_2+\mu_1 &\le 3\nu_1+\nu_2+\nu_3-\nu_6-\nu_4-3\nu_5,\\
2\lambda_3-2\lambda_1-\mu_2+\mu_1 &\le 3\nu_1+\nu_2+\nu_5-\nu_3-\nu_6-3\nu_4,\\
2\lambda_3-2\lambda_1+\mu_2-\mu_1 &\le
3\nu_1+\nu_2+\nu_6-\nu_3-\nu_5-3\nu_4.
\displaybreak[0]\\[2mm]
2\lambda_1+2\lambda_2-4\lambda_3+3\mu_1-3\mu_2 &\le
5\nu_1+5\nu_2-\nu_3-\nu_4-\nu_5-7\nu_6,\\
2\lambda_3+2\lambda_1-4\lambda_2+3\mu_1-3\mu_2 &\le
5\nu_3+5\nu_1-\nu_2-\nu_4-\nu_5-7\nu_6,\\
2\lambda_1+2\lambda_3-4\lambda_2+3\mu_2-3\mu_1 &\le
5\nu_2+5\nu_3-\nu_1-\nu_4-\nu_5-7\nu_6,\\
2\lambda_2+2\lambda_3-4\lambda_1+3\mu_1-3\mu_2 &\le
5\nu_2+5\nu_3-\nu_1-\nu_4-\nu_5-7\nu_6,\\
2\lambda_2+2\lambda_1-4\lambda_3+3\mu_2-3\mu_1 &\le
5\nu_1+5\nu_2-\nu_3-\nu_6-\nu_4-7\nu_5,\\
2\lambda_3+2\lambda_1-4\lambda_2+3\mu_1-3\mu_2 &\le
5\nu_1+5\nu_2-\nu_3-\nu_6-\nu_4-7\nu_5,\\
2\lambda_2+2\lambda_3-4\lambda_1+3\mu_1-3\mu_2 &\le
5\nu_1+5\nu_3-\nu_2-\nu_6-\nu_4-7\nu_5,\\
2\lambda_2+2\lambda_3-4\lambda_1+3\mu_1-3\mu_2 &\le
5\nu_1+5\nu_2-\nu_3-\nu_5-\nu_6-7\nu_4.
\displaybreak[0]\\[2mm]
4\lambda_1-2\lambda_2-2\lambda_3+3\mu_1-3\mu_2 &\le
7\nu_1+\nu_2+\nu_3+\nu_4-5\nu_5-5\nu_6,\\
4\lambda_1-2\lambda_2-2\lambda_3+3\mu_2-3\mu_1 &\le
7\nu_2+\nu_1+\nu_3+\nu_4-5\nu_5-5\nu_6,\\
4\lambda_2-2\lambda_1-2\lambda_3+3\mu_1-3\mu_2 &\le
7\nu_2+\nu_1+\nu_3+\nu_4-5\nu_5-5\nu_6,\\
4\lambda_2-2\lambda_1-2\lambda_3+3\mu_1-3\mu_2 &\le
7\nu_1+\nu_2+\nu_3+\nu_5-5\nu_4-5\nu_6,\\
4\lambda_3-2\lambda_1-2\lambda_2+3\mu_1-3\mu_2 &\le
7\nu_3+\nu_1+\nu_2+\nu_4-5\nu_5-5\nu_6,\\
4\lambda_3-2\lambda_1-2\lambda_2+3\mu_1-3\mu_2 &\le
7\nu_2+\nu_1+\nu_3+\nu_5-5\nu_4-5\nu_6,\\
4\lambda_2-2\lambda_1-2\lambda_3+3\mu_2-3\mu_1 &\le
7\nu_1+\nu_2+\nu_3+\nu_6-5\nu_4-5\nu_5,\\
4\lambda_3-2\lambda_1-2\lambda_2+3\mu_1-3\mu_2 &\le
7\nu_1+\nu_2+\nu_3+\nu_6-5\nu_4-5\nu_5.
\end{align*}}

\subsubsection{Format $2\times4$}\label{2x4}
Marginal constraints in this case are given by the following 234
independent inequalities. Note that three qubit case can be
reduced to this one.
{\footnotesize
}

\subsection{System of format $2\times2\times3$}\label{2x2x3}
In contrast with other systems of rank $\le4$ in this case there
are $9$ {\it redundant extremal edges\,} for which all the
associated inequalities are redundant.
For all the other systems basic inequalities are essential and
independent.

As usual we skip inequalities obtained from another one by a
permutation of qubits. This reduces the number of marginal
inequalities to $232$ (instead of $442$).
Under additional constraint $\lambda_1-\lambda_2\ge
\mu_1-\mu_2$ they form a complete and independent system. {\footnotesize
\begin{align*}
\lambda_1-\lambda_2 &\le
\rho_1+\rho_2+\rho_3+\rho_4+\rho_5+\rho_6-\rho_7-\rho_8-\rho_9-\rho_{10}-\rho_{11}-\rho_{12}.
\displaybreak[0]\\[2mm]
2\nu_1-\nu_2-\nu_3 &\le
2\rho_1+2\rho_2+2\rho_3+2\rho_4-\rho_5-\rho_6-\rho_7-\rho_8-\rho_9-\rho_{10}-\rho_{11}-\rho_{12}.
\displaybreak[0]\\[2mm]
\nu_1+\nu_2-2\nu_3 &\le
\rho_1+\rho_2+\rho_3+\rho_4+\rho_5+\rho_6+\rho_7+\rho_8-2\rho_9-2\rho_{10}-2\rho_{11}-2\rho_{12}.
\displaybreak[0]\\[2mm]
\lambda_1-\lambda_2+\mu_1-\mu_2 &\le
2\rho_1+2\rho_2+2\rho_3-2\rho_{10}-2\rho_{11}-2\rho_{12}.
\displaybreak[0]\\[2mm]
\lambda_1-\lambda_2+\nu_1-\nu_3 &\le
2\rho_1+2\rho_2+\rho_3+\rho_4-\rho_9-\rho_{10}-2\rho_{11}-2\rho_{12}.
\displaybreak[0]\\[2mm]
3\lambda_1-3\lambda_2+2\nu_1-\nu_2-\nu_3 &\le
5\rho_1+5\rho_2+2\rho_3+2\rho_4+2\rho_5+2\rho_6-\rho_7-\rho_8-4\rho_9-4\rho_{10}-4\rho_{11}-4\rho_{12},\\
3\lambda_1-3\lambda_2+2\nu_3-\nu_1-\nu_2 &\le
5\rho_2+5\rho_3+2\rho_1+2\rho_4+2\rho_5+2\rho_6-\rho_7-\rho_8-4\rho_9-4\rho_{10}-4\rho_{11}-4\rho_{12},\\
3\lambda_1-3\lambda_2+2\nu_3-\nu_1-\nu_2 &\le
5\rho_1+5\rho_2+2\rho_3+2\rho_4+2\rho_5+2\rho_6-\rho_8-\rho_9-4\rho_7-4\rho_{10}-4\rho_{11}-4\rho_{12}.
\displaybreak[0]\\[2mm]
3\lambda_1-3\lambda_2+\nu_1+\nu_2-2\nu_3 &\le
4\rho_1+4\rho_2+4\rho_3+4\rho_4+\rho_5+\rho_6-2\rho_7-2\rho_8-2\rho_9-2\rho_{10}-5\rho_{11}-5\rho_{12},\\
3\lambda_1-3\lambda_2+\nu_2+\nu_3-2\nu_1 &\le
4\rho_1+4\rho_2+4\rho_3+4\rho_6+\rho_4+\rho_5-2\rho_7-2\rho_8-2\rho_9-2\rho_{10}-5\rho_{11}-5\rho_{12},\\
3\lambda_1-3\lambda_2+\nu_2+\nu_3-2\nu_1 &\le
4\rho_1+4\rho_2+4\rho_3+4\rho_4+\rho_5+\rho_6-2\rho_7-2\rho_8-2\rho_9-2\rho_{12}-5\rho_{10}-5\rho_{11}.
\end{align*}
}



\bibliographystyle{amsplain}
\bibliography{Q_Margins}
\end{document}